\documentclass[journal]{IEEEtran}
\usepackage{cite}
\ifCLASSINFOpdf
  \usepackage[pdftex]{graphicx}
\else
  \usepackage[dvips]{graphicx}
\fi
\usepackage{amssymb}
\usepackage{algorithm}
\usepackage{algorithmic}
\usepackage{tabularx}
\usepackage{amsthm,amssymb,amsfonts}
\usepackage{epsfig}
\usepackage{subfigure}
\usepackage{array}
\usepackage{changes}
\usepackage{textcomp}
\usepackage{xcolor}
\usepackage{amsmath}
\usepackage{fancyhdr}
\usepackage{makecell}
\usepackage{dsfont}
\usepackage{amsmath,amssymb,amsfonts}
\usepackage{subfigure}
\usepackage{textcomp}
\usepackage{xcolor}
\usepackage{multirow}
\usepackage{mathtools}
\usepackage{mhchem}
\usepackage{tensor}
\usepackage{graphicx}  
\usepackage{psfrag}    
\usepackage{amsthm}
\usepackage{mathrsfs}
\usepackage{epstopdf}
\usepackage{bm}
\usepackage{cuted}
\usepackage{color, soul, framed}

\usepackage{xcolor}

\usepackage{xpatch}
\usepackage{tabularx}
\ifCLASSOPTIONcompsoc
  \usepackage[caption=false,font=normalsize,labelfont=sf,textfont=sf]{subfig}
\else
  \usepackage[caption=false,font=footnotesize]{subfig}
\fi
\usepackage{fixltx2e}
\usepackage{dblfloatfix}

\usepackage{multirow}
\usepackage{booktabs}
\newcommand{\tabincell}[2]{\begin{tabular}{@{}#1@{}}#2\end{tabular}}
\usepackage{url}
\begin{document}

\title{Generalized code index modulation-aided frequency offset realign multiple-antenna spatial modulation approach for next-generation green communication systems}

\author{Bang Huang,~\IEEEmembership{Member,~IEEE},	Jiajie Xu,~\IEEEmembership{Member,~IEEE},
	Mohamed-Slim Alouini,~\IEEEmembership{Fellow,~IEEE}
		
	\thanks{ The authors are with the Computer, Electrical, and Mathematical Science and Engineering (CEMSE) division in King Abdullah University of Science and Technology (KAUST), Thuwal 6900, Makkah Province, Saudi Arabia.(Emails: bang.huang@kaust.edu.sa; jiajie.xu.1@kaust.edu.sa; slim.alouini@kaust.edu.sa) (Corresponding author: Jiajie Xu)}}
\maketitle

\begin{abstract}
For next-generation green communication systems, this article proposes an innovative communication system based on frequency-diverse array-multiple-input multiple-output (FDA-MIMO) technology, which aims to achieve high data rates while maintaining low power consumption. This system utilizes frequency offset index realign modulation, multiple-antenna spatial index modulation, and spreading code index modulation techniques. In the proposed generalized code index modulation-aided frequency offset realign multiple-antenna spatial modulation  (GCIM-FORMASM) system, the coming bits are divided into five parts: spatial modulation bits by activating multiple transmit antennas, frequency offset index bits of the FDA antennas, including frequency offset combination bits and frequency offset realign bits, spreading code index modulation bits, and modulated symbol bits. Subsequently, this paper utilizes the orthogonal waveforms transmitted by the FDA to design the corresponding transmitter and receiver structures and provide specific expressions for the received signals. Meanwhile, to reduce the decoding complexity of the maximum likelihood (ML) algorithm, we propose a three-stage despreading-based low complexity (DBLC) algorithm leveraging the orthogonality of the spreading codes. Additionally, a closed-form expression for the upper bound of the average bit error probability (ABEP) of the DBLC algorithm has been derived. Analyzing metrics such as energy efficiency and data rate shows that the proposed system features low power consumption and high data transmission rates, which aligns better with the concept of future green communications. 
The effectiveness of our proposed methods has been validated through comprehensive numerical results.
\end{abstract}


\IEEEpeerreviewmaketitle

\section{Introduction}

\IEEEPARstart{T}{he}
focus of fifth-generation (5G) mobile communication gradually shifted from improving call quality to applications that demand high transmission rates, such as large file sharing and high-quality video \cite{duan2020emerging}. 
However, the substantial number of users and base stations in 5G wireless networks significantly increase energy consumption \cite{gandotra2017survey}. Therefore, designing a method with high transmission rates and low power consumption for next-generation communication systems is an urgent challenge that needs to be addressed \cite{buzzi2016survey}.

It is foreseeable that both 5G and beyond communications systems will extensively use multiple-input multiple-output (MIMO) antenna technology \cite{li2010mimo,wen2017index}. MIMO, by employing multiple antennas at both the transmitter and receiver ends, can effectively enhance system data rate, frequency spectral, and error performance on the one hand and explore multi-channel capabilities to boost channel capacity and diversity gain on the other\cite{mietzner2009multiple}. In recent years, the hottest and most significant research direction among all MIMO technologies has been the index modulation (IM) technique due to its potential to balance energy efficiency required by the beyond green communications \cite{rathore2021green} and spectral efficiency \cite{basar2016index}. IM technology typically requires very little or no additional energy to embed extra bits into the transmitted signal through modulating indexes of units, for example, the transmit antennas \cite{mesleh2008spatial},  spreading codes \cite{kaddoum2014code}, subcarriers or frequency carriers \cite{bacsar2013orthogonal,Huang2020IndexModulation},  and so on \cite{mao2018novel}. 

As an essential component of IM technology, the introduction and exploration of spatial modulation (SM)  have significantly advanced the development of IM \cite{wen2019survey}. In the SM framework, source bits are divided into modulation symbols, and additional spatial index bits are obtained by activating a single antenna within a time slot. Since SM activates only one antenna, it effectively addresses inter-channel interference (ICI) and synchronization issues between antennas, significantly reducing the complexity of the receiver \cite{yang2014design}. 
 However, in the SM method, increasing one data bit transmitted requires doubling the number of transmit antennas, which is very uneconomical. To address this issue, a generalized SM (GSM) method has been proposed \cite{wang2012generalised}, where multiple transmit antennas can be activated in each time slot. Additionally, quadrature SM (QSM) aims to double the transmitted spatial index bits by extending the spatial constellation to both in-phase and quadrature dimensions \cite{mesleh2014quadrature}. 

Another method to achieve IM modulation is code index modulation spread spectrum (CIM-SS) technology by applying direct sequence spread spectrum (DS-SS) \cite{kaddoum2014code}. Via the indices of orthogonal spreading code, the additional information bits have been transmitted in CIM-SS system, which has been endowed the advantages of high data rates, high energy efficient. Moreover, \cite{kaddoum2015generalized} proposed a generalized CIM (GCIM) technique by modulating the real and imaginary parts of the symbol separately using spreading codes. 
Next, {\c{C}}{\"o}gen, \textit{et al.} \cite{ccogen2018novel} were the first to propose the integration of CIM and SM technologies, namely CIM-SM.
Compared to CIM and SM modulation, this approach offers higher data rates, increased energy efficiency, and lower bit error rates (BER). However, CIM-SM is limited to binary phase shift keying (BPSK) symbol modulation, which restricts the transmission to only 1 bit per modulated symbol per time slot. To overcome this limitation, \cite{cogen2020generalized} introduces a generalized CIM-SM (GCIM-SM) modulation technique that extends the constellation modulation from BPSK to
$M$-ary quadrature amplitude modulation (QAM) symbol modulation. This GCIM-SM approach surpasses CIM, QSM, SM, and CIM-SM technologies by achieving high data rates, high spectral efficiency, and high energy efficiency with minimal hardware modifications. 
On the other hand, combining CIM with QSM to form CIM-QSM technology \cite{aydin2019code} with enhancing spectral efficiency and BER performance. In the CIM-aided SM and media-based modulation (MBM) (CIM-SMBM) system, information bits can be conveyed not only through symbol mapping but also by activating SM-selected active antennas, selecting different channel state indices via MBM and determining spreading code indices via CIM. To further improve spectrum efficiency and data rate, CIM technology is integrated with differential chaos shift keying (DCSK) \cite{xu2018design}, code-shifted DCSK (CS-DCSK) \cite{xu2017code}, and multi-carrier $M$-ary DCSK (MC-M-DCSK) \cite{cai2019multi},
 respectively.

Another important branch of IM is extending the index dimension to the frequency domain, including subcarriers and multiple transmit carriers. Subcarrier indexing primarily refers to combining orthogonal frequency division multiplexing (OFDM) and IM technology to form the OFDM-IM scheme. Specifically, in their pioneering work, Ba{\c{s}}ar \textit{et al.} \cite{bacsar2013orthogonal} formed a pool of subcarriers and then selected a subset of subcarriers from the pool in each time slot as needed to transmit together with constellation symbols. 
Furthermore, Fan \textit{et al.} introduced two generalized OFDM-IM (OFDM-GIM) schemes in \cite{fan2015generalization} to enhance spectral efficiency.
Motivated by this, \cite{wen2017enhanced} proposed two enhanced OFDM-IM schemes to achieve higher spectral efficiency and diversity gain. In-depth discussions on OFDM-IM technology can also be found in \cite{xu2022orthogonal,li2022composite} and their associated literature lists.

Although \cite{cai2019multi} transmits information-carrying signals across different carrier frequencies, the multiple carriers do not convey additional bit symbols. Unlike \cite{cai2019multi}, Huang \textit{et al.} \cite{Huang2020IndexModulation} applied IM technology to the frequency diverse array (FDA), an emerging area that has garnered significant attention in the radar field. Compared to OFDM, FDA distributes different subcarriers across various array elements \cite{HuangYan2022RadarCrossSection}, leading to frequency offsets among different elements \cite{JianWang2023PhysicalLayerSecurity}. Consequently, IM-FDA can transmit information by indexing different frequency offsets. 
Subsequently, \cite{Nusenu2020SpaceFrequency} proposed a space-frequency increment index modulation method based on the principles of QSM for the FDA. This method conveyed additional bit information by activating different FDA antennas and their corresponding frequency offsets. 
Moreover, \cite{jian2023mimo} combined FDA with IM technology to introduce the frequency offsets index modulation (FOIM) system. This system transmits additional index modulation information by selecting different transmission frequency offsets from a frequency offset pool without rearranging the transmitted frequency offsets. In \cite{jian2023fda}, the same authors addressed this limitation by rearranging the selected frequency offsets, which not only improved the communication rate but also investigated the system's performance on the radar end. However, neither FOIM nor FOPIM leverages the transmission antenna index or the spreading code index. This has motivated us to explore these aspects further. More importantly, it can be seen from the preceding parts that index modulation employed more dimensions significantly increases data rates, improves spectrum efficiency, and conserves energy.

Inspired by the aforementioned scheme, this paper introduces a novel system called generalized code index modulation-aided frequency offset realign multiple-antenna spatial modulation (GCIM-FORMASM) based on FDA-MIMO technology. This system is designed to enhance data rate, spectral efficiency, and energy efficiency for next-generation green communication. The key contributions of this paper are summarized as follows:
\begin{itemize}
	\item \textbf{Innovative System Design}: We propose the GCIM-FORMASM system, which integrates GCIM with frequency offset realign index modulation and multiple-antenna spatial modulation. Unlike existing CIM-based SM research, our approach incorporates the FDA to significantly enhance the data rate. Furthermore, our comparison with FOIM and FOPIM systems demonstrates that we are the first to introduce additional GCIM frameworks by activating more than one antenna or all antennas (it means the GSM technology), which not only improves spectral efficiency but also reduces the number of active antennas, thereby enhancing energy efficiency.
	\item \textbf{Low-Complexity Algorithm Development}: This paper proposes a high-complexity ML detector for the GCIM-FORMASM system. In addition, to further reduce the complexity of the ML detector, we design a three-step despreading-based low-complexity (DBLC) detector. Specifically, the selected frequency offsets are estimated by searching for the maximum output values at the receiver. Secondly, despreading is performed on the spreading codes to demodulate the spreading code, the activated transmit antennas, and the frequency offset realignment. Finally, the ML criterion is used to estimate the QAM-modulated symbols transmitted by each antenna corresponding to the frequency offset and spreading code.
	\item \textbf{Analytical Performance Evaluation}: We derive a closed-form expression for the upper bound of the average bit error probability (ABEP) for the DBLC algorithm. Analyzing metrics such as energy efficiency and data rate indicate that compared to existing systems such as FOPIM, GSCIM, GCIM-SM, and SM, we demonstrate that the proposed system achieves low power consumption, aligning with the principles of future green communication.
	\item  \textbf{Extensive Simulation Validation}: Numerical simulation results validate the effectiveness of the proposed DBLC algorithm, showing superior BER performance compared to existing algorithms, both in fair comparisons (transmitting the same number of bits) and unfair comparisons (different information bits under the same system framework). Additionally, the DBLC algorithm exhibits lower computational complexity and better BER performance compared to the ML algorithm.
\end{itemize}

The structure of the paper is as follows. Section \ref{sec2} details the formulation of the GCIM-FORMASM system model, including both the transmitter and receiver diagrams. In Section \ref{sec3}, we conduct a comprehensive performance analysis of the proposed system. Section \ref{sec4} presents the simulation results, while Section \ref{sec5} concludes the paper, summarizing the essential findings and implications of our work.

\textit{Notations:} The matrices (vectors) are represented by boldface upper (lower) case letters. 
$\mathbb{C}^{m\times n}$ represents the complex matrix space of dimension $m\times n$. 
The symbol $
\lfloor \cdot \rfloor 
$ represents the floor function, which rounds a number down to the nearest integer. Additionally, $
C_{M}^{N}
$ is the combination of choosing N elements from M and $N!$ is the factorial of N. The symbol $
\mathrm{sort}\left\{ \cdot \right\} 
$ represents arranging the elements of the set in descending order. The symbol $
\mathrm{rem}\left( \mathrm{x},\mathrm{y} \right) 
$ represents the remainder when $\mathrm{x}$ is divided by $\mathrm{y}$ and meanwhile $\mathrm{mod}\left( \mathrm{x},\mathrm{y} \right)$ represents the integer quotient of $\mathrm{x}$ divided by $\mathrm{y}$.

\section{System Model}
\label{sec2}

This section will provide more details about the transmitter and receiver structures for the proposed GCIM-FORMASM system. Besides, the assumption of estimating the perfect channel can be obtained for the receiver.
\begin{figure}[htp]
	\centering
	{\includegraphics[width=0.45\textwidth]{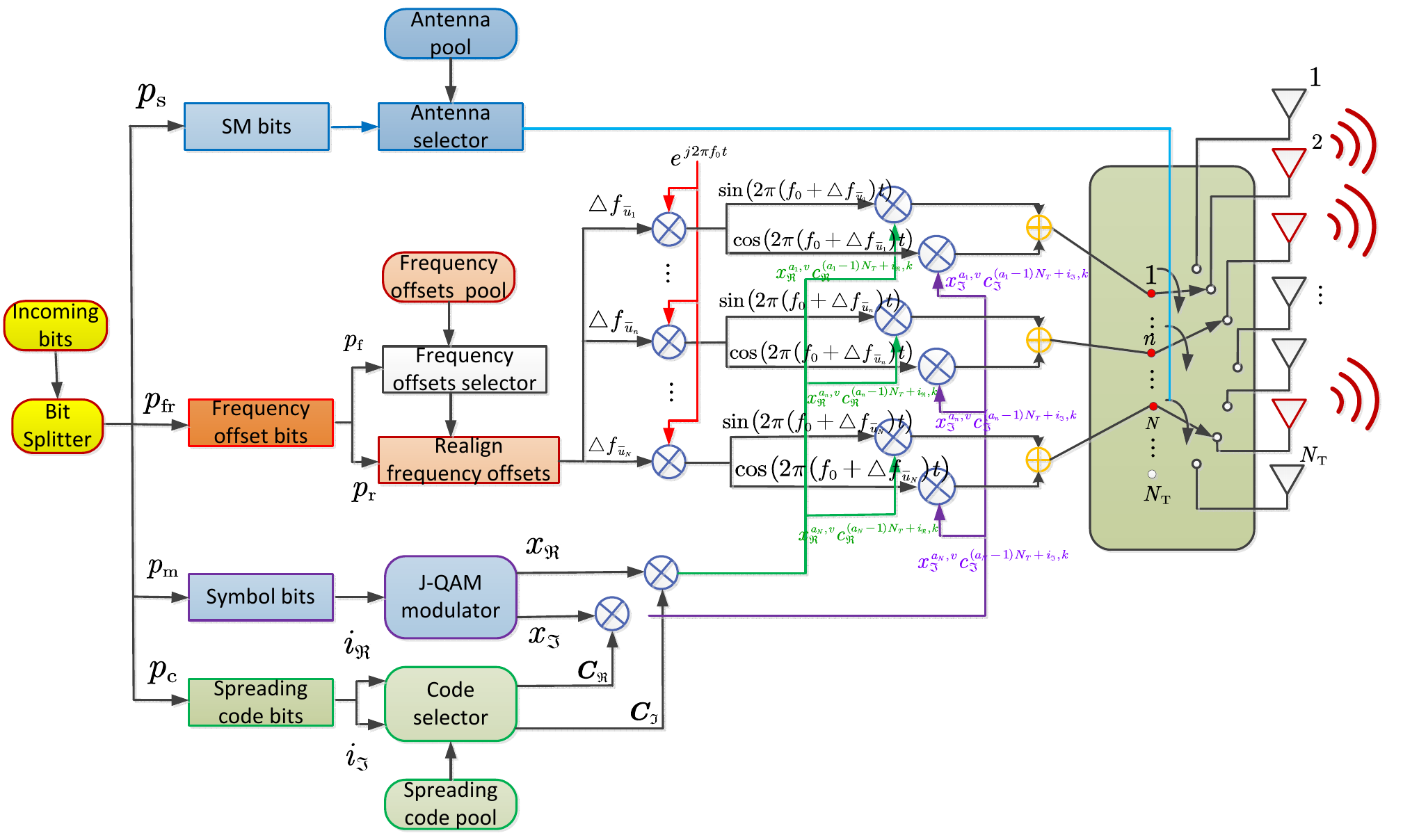}}
	\caption{The transmitter structure for our proposed GCIM-FORMASM scheme.}
	\label{fig1}	
\end{figure}
\subsection{Transmitter diagram}  
Fig.\ref{fig1} presents the transmitter diagram for the proposed GCIM-FORMASM system. This structure includes an FDA-based transmit uniform linear array (ULA) with $N_T$ antennas. To reduce the energy consumption, there are  $N,
\left( 1< N\leqslant N_T \right) 
$ active antennas during transmitting the message, which means that the transmitted bits in the antenna indices of the SM can be expressed as $
p_s=\lfloor \log _2C_{N_T}^{N} \rfloor 
$. It is noteworthy that the part of FDA-based frequency offset realign index modulation is the same as the method in \cite{jian2023fda}, which indicates that we can directly use the results to enhance the system's performance. Hence, the frequency offset bits $
p_{fr}
$ can be further divided into two parts, namely frequency offset combination bits $
p_{f}
$ and frequency offset permutation bits $
p_{r}
$. Assume there exists a frequency offset pool $
\mathcal{M} =\{ \bigtriangleup f^1,\bigtriangleup f^2,\cdots ,\bigtriangleup f^{ M } \} 
\in \mathbb{C} ^M
$, where $
\bigtriangleup f^{m_1}<\bigtriangleup f^{m_2}
$ for any ${m_1}<m_2$ with $
1\leqslant m_1,m_2\leqslant M
$\footnote{We call the linear frequency increment when $
	\bigtriangleup f^m=m\bigtriangleup f
	$.
	 If not, it is the non-linear frequency offset across the adjacent elements. 
   Interesting readers may refer \cite{Wang2015FrequencyDiverseArray,choi2022analysis} to obtain more details.
  }, as well as the following orthogonal relationship $	 \int_{T_c}{e^{j2\pi \left( \bigtriangleup f^{m_1}-\bigtriangleup f^{m_2} \right) t}}dt=\left\{ \begin{array}{c}
	 	0,m_1=m_2\\
	 	T_c,m_1\ne m_2\\
	 \end{array} \right. $ 
with $T_c$ is the chip duration.
	  Then we have $
C_{M}^{N}
$ choices to select the frequency offsets from the above pool to carry $
p_f=\lfloor \log _2C_{M}^{N} \rfloor 
$ information bits. Next, the choosing $N$ frequency increments can be realigned to transmit $
p_r=\lfloor \log _2N! \rfloor 
$ bits. 
On the other hand, $
p_m=N\log _2J
$ denotes the constellation bits for $N$ unit power $J$-ary QAM symbols. Also, our proposed system includes $
L
$ Walsh Hadamard codes for $I$ or $Q$ components. In this respect, there are $
p_c=2N\log _2L
$ spreading code bits map for the $I/Q$ components. Therefore, the carrying bits for 
the proposed GCIM-FORMASM system can be given as $p=p_s+p_f+p_r+p_m+p_c$.

As evident in Fig.\ref{fig1}, all coding symbol sequences are transmitted through the activation of $M$ antennas under the SM scheme.
Compared with the \cite{cogen2020generalized,cogen2018CodeIndexModulation}, our proposed diagram activates more than one antenna and takes the FDA into consideration to highly improve the data rate. Moreover, the comparison between this paper and \cite{jian2023fda,jian2023mimo} implies that we first provide additional GCIM and GSM frameworks to boost the spectral efficiency as well as enhance the energy efficiency.

When there are $N$ antennas from the $N_T$ FDA antennas to transmit the message, the corresponding activating antenna pool can be expressed as $
\mathbb{A} 
=
\left\{ a_1,\cdots ,a_n,\cdots a_N \right\} 
$ with $
a_n\in \left\{ 1,\cdots ,N_T \right\} 
$ denoting the location of the activating antenna. Additionally, for any $
n_1<n_2
$ with $
n_1,n_2\in \left\{ 1,\cdots ,N \right\}  
$, we have $
a_{n_1}<a_{n_2}
$. Subsequently, the selected $N$ frequency increments from the frequency offsets pool can be grouped into the following set: $
\tilde{\mathcal{M}}=\left\{ \bigtriangleup f_{u_1},\cdots ,\bigtriangleup f_{u_n},\cdots ,\bigtriangleup f_{u_N} \right\} 
$ with $
\left\{ u_1,\cdots ,u_n,\cdots u_N \right\} \in \mathbb{Z} ^N
$, $
u_n\in \left\{ 1,\cdots ,M \right\} 
$ representing the location of involved frequency offset in $\mathcal{M}$. In this case, we have $
u_1<\cdots <u_n<\cdots <u_N
$.

Realigning the chosen frequency offsets pool results in the new pool  $
\tilde{\mathcal{M}}_1=\left\{ \bigtriangleup f_{\bar{u}_1},\cdots ,\bigtriangleup f_{\bar{u}_n},\cdots ,\bigtriangleup f_{\bar{u}_N} \right\} 
$ with $
\left\{ \bar{u}_1,\cdots ,\bar{u}_n,\cdots ,\bar{u}_N \right\} \in \mathbb{Z} ^N
$ and $
\bar{u}_n\in \left\{ u_1,\cdots ,u_n,\cdots u_N \right\} 
$. Further, the modulated $J$-ary QAM pool can be written as 
\begin{equation}
	\boldsymbol{X}=\left\{ \boldsymbol{x}_1,\cdots ,\boldsymbol{x}_n,\cdots ,\boldsymbol{x}_{N_T} \right\} \in \mathbb{C} ^{J\times N_T},\nonumber
\end{equation}with 
$\boldsymbol{x}_n=[ x_{}^{n,1},\cdots ,x_{}^{n,v},\cdots ,x_{}^{n,J} ] ^T\in \mathbb{C} ^{J\times 1}$,
which indicates the modulated information transmitted by the $n$th antenna as well as the message $x_{}^{n,v}$ can be divided into the real and imaginary parts, as $
x_{}^{n,v}=x_{\mathfrak{R}}^{n,v}+jx_{\mathfrak{I}}^{n,v}
$ with $v={1,2,...,2^{p_m}}$. Finally, assume there are two code pools, like 
\begin{align}
	\boldsymbol{C}_{\Re}=&\{ \boldsymbol{C}_{\Re}^{1},\cdots ,\boldsymbol{C}_{\Re}^{n},\cdots ,\boldsymbol{C}_{\Re}^{N_T} \} \in \mathbb{R} ^{K\times LN_T},\nonumber
	\\
	\boldsymbol{C}_{\mathfrak{I}}=&\{ \boldsymbol{C}_{\mathfrak{I}}^{1},\cdots ,\boldsymbol{C}_{\mathfrak{I}}^{n},\cdots ,\boldsymbol{C}_{\mathfrak{I}}^{N_T} \} \in \mathbb{R} ^{K\times LN_T},\nonumber
\end{align}
where the spreading Walsh codes transmitting by the $n$th antenna through the $I$ and $Q$ components are

\begin{align}
	\boldsymbol{C}_{\Re}^{n}=&[ \boldsymbol{c}_{\Re}^{n,1},\cdots ,\boldsymbol{c}_{\Re}^{n,i_{\Re}},\cdots ,\boldsymbol{c}_{\Re}^{n,L} ] \in \mathbb{R} ^{K\times L},\nonumber
	\\
	\boldsymbol{C}_{\mathfrak{I}}^{n}=&[ \boldsymbol{c}_{\mathfrak{I}}^{n,1},\cdots ,\boldsymbol{c}_{\mathfrak{I}}^{n,i_{\mathfrak{I}}},\cdots ,\boldsymbol{c}_{\mathfrak{I}}^{n,L} ] \in \mathbb{R} ^{K\times L},\nonumber
\end{align} 
with 
\begin{align}
	\boldsymbol{c}_{\Re}^{n,i_{\Re}}=&\big[ c_{\Re}^{\left( n-1 \right) L+i_{\Re},1},\cdots ,c_{\Re}^{\left( n-1 \right) L+i_{\Re},K} \big] ^T\in \mathbb{C} ^K,\nonumber
\\
\boldsymbol{c}_{\mathfrak{I}}^{n,i_{\mathfrak{I}}}=&\big[ c_{\mathfrak{I}}^{\left( n-1 \right) L+i_{\Re},1},\cdots ,c_{\mathfrak{I}}^{\left( n-1 \right) L+i_{\Re},K} \big] ^T\in \mathbb{C} ^K.\nonumber
\end{align}
The symbol $K$ is the total transmitted chips and $
i_{\Re},i_{\mathfrak{I}}=1,2,\cdots ,L
$.

Moreover, the expression for the signal for $a_n$th transmit antenna at the $k$th chip is 
\begin{align}
	\label{eq10}
y_{a_n,k}( t ) =&\sqrt{\frac{P_S}{N}}p( t-kT_c ) \Big[ \cos [ 2\pi ( f_0+\bigtriangleup f_{\bar{u}_n,a_n}^{} ) ( t-\tau _{a_n} ) ]  \nonumber
\\
&\times x_{\Re}^{n,v}c_{\Re}^{( a_n-1 ) N_T+i_{\Re},k}+j\sin [ 2\pi ( f_0+\bigtriangleup f_{\bar{u}_n,a_n}^{} )  \nonumber
\\
&\times \left. \left( t-\tau _{a_n} \right) \right] x_{\Im}^{n,v}c_{\Im}^{\left( a_n-1 \right) N_T+i_{\Im},k} \Big], 
\end{align}
where $p\left( t \right)$ and $P_S$ denoting the unit rectangular pulse shaping filter on $
\left[ 0,T_c \right] 
$ time period and the transmitted power, respectively.

\subsection{Receiver diagram} 
This paper considers a Rayleigh fading MIMO channel, meaning that it is selective to frequency \cite{bacsar2013orthogonal}. Hence, the channel cofficient matrix is given by $	\boldsymbol{H}=\left[ \boldsymbol{H}_1,\cdots ,\boldsymbol{H}_m,\cdots ,\boldsymbol{H}_M \right] \in \mathbb{C} ^{N_R\times N_TM},$
where $N_R$ denotes the number of receiving antenna and the matrix $\boldsymbol{H}_m$ can be expressed as $	\boldsymbol{H}_m=\left[ \boldsymbol{h}_{m,1},\cdots ,\boldsymbol{h}_{m,n},\cdots ,\boldsymbol{h}_{m,N_T} \right] \in \mathbb{C} ^{N_R\times N_T},$
with $\boldsymbol{h}_{m,n}=[ h_{\left( m-1 \right) N_T+n,1},\cdots ,h_{\left( m-1 \right) N_T+n,N_R} ] ^T\in \mathbb{C} ^{N_R}.$
Note that each column of $\boldsymbol{H}$ follows an independent identically distributed (I.I.D) Gaussian distribution, i.e., $
\boldsymbol{h}_{m,n}\sim \mathcal{C} \mathcal{N} ( 0,\sigma ^2\boldsymbol{I} ) 
$.

Further, the receive signal model for $n_r$th antenna at $k$th chip can be written as
\begin{align}
	\label{eq14}
y_{n_r,k}( t ) =&\sqrt{\frac{P_S}{N}}\sum_{a_n=a_1}^{a_N}{p( t-kT_c ) \Big[ \cos [ 2\pi ( f_0+\bigtriangleup f_{\bar{u}_n,a_n}^{} ) }\nonumber
\\
&\times ( t-{\tau _{a_n}}_{,n_r} ) x_{\Re}^{a_n,v}c_{\Re}^{( a_n-1 ) N_T+i_{\Re},k}\nonumber
\\
&+j\sin [ 2\pi ( f_0+\bigtriangleup f_{u_{\bar{n}},a_n}^{} ) ( t-{\tau _{a_n}}_{,n_r} ) ] \nonumber
\\
&\times  x_{\Im}^{a_n,v}c_{\Im}^{( a_n-1 ) N_T+i_{\Im},k} \Big] h_{( \bar{u}_n-1 ) N_T+a_n,n_r},
\end{align}
where $
{\tau _{a_n}}_{,n_r}=\frac{r+a_nd\sin \theta +n_rd\sin \theta}{c}
$ representing the time delay.

\begin{figure}[htp]
	\centering
	{\includegraphics[width=0.48\textwidth]{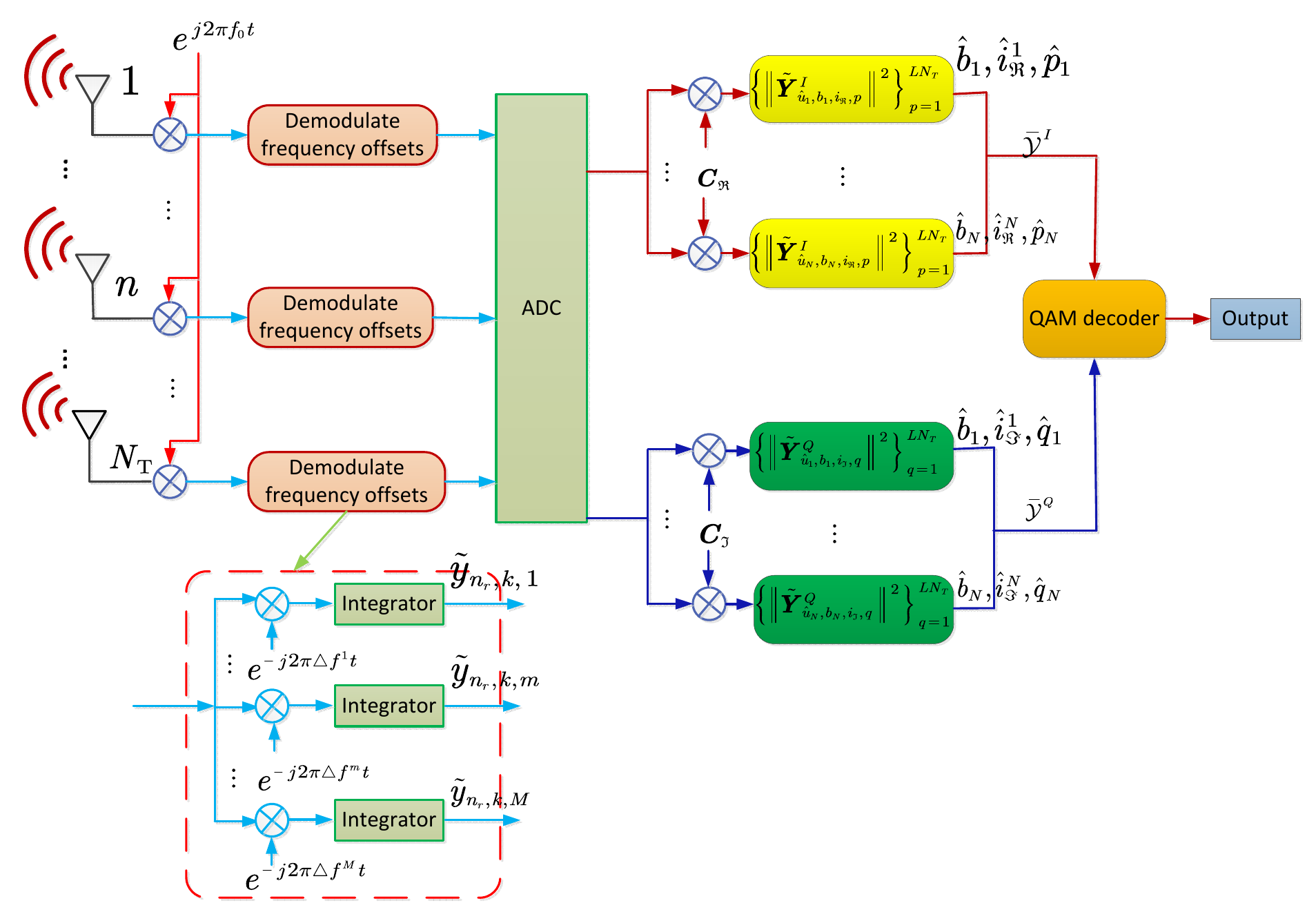}}
	\caption{The receiver diagram for our proposed GCIM-FORMASM system.}
	\label{fig2}	
\end{figure}

We propose a receiver architecture designed to efficiently demodulate the signals transmitted by our GCIM-FORMASM scheme, depicted in Fig.\ref{fig2}. Specifically,
removing the carrier frequency in \eqref{eq14}  gives 
\begin{align}
			\tilde{y}_{n_r,k}( t ) =&\sqrt{\frac{P_S}{N}}\sum_{a_n=a_1}^{a_N}{p( t-kT_c ) \Big[ \cos [ -2\pi f_0{\tau _{a_n}}_{,n_r} }\nonumber
		\\
		& +2\pi \bigtriangleup f_{\bar{u}_n,a_n}^{}( t-{\tau _{a_n}}_{,n_r} ) ] x_{\Re}^{a_n,v}c_{\Re}^{( a_n-1 ) N_T+i_{\Re},k}\nonumber
		\\
		&+j\sin [ -2\pi f_0{\tau _{a_n}}_{,n_r}+2\pi \bigtriangleup f_{\bar{u}_n,a_n}^{}( t-{\tau _{a_n}}_{,n_r} ) ] \nonumber
		\\
		&\times  x_{\Im}^{a_n,v}c_{\Im}^{( a_n-1 ) N_T+i_{\Im},k} \Big] h_{( \bar{u}_n-1 ) N_T+a_n,n_r}.
\end{align} 

Subsequently, using the orthogonal relationship mentioned above yields the output signal  in $m$th, $
m=\left\{ 1,2,\cdots ,M \right\} 
$,
 channel located at the $n_r$th frequency offset decoder block with noise, expressed in \eqref{eq17} (shown at the top of next page), where $
 v_{\left( n_r-1 \right) M+m,k}
 $ denotes the additive Gaussian white noise (AWGN) in the $m$th channel of $n_r$th receive for the $k$th chip, namely $
 v_{\left( n_r-1 \right) M+m,k}\sim \mathcal{C} \mathcal{N} \left( 0,{N_0}{} \right) 
 $ with $N_0$ being the total noise power. 
\begin{figure*}
	\begin{align}
		\label{eq17}
			\tilde{y}_{n_r,k,m}=&\int{\tilde{y}_{n_r,k}\left( t \right)}e^{j2\pi \bigtriangleup f^mt}dt\nonumber
			\\
			=&\Bigg\{ \begin{array}{r}\!\!
				\sqrt{\frac{P_S}{N}}\big[ x_{\Re}^{a_n,v}c_{\Re}^{\left( a_n-1 \right) N_T+i_{\Re},k}+jx_{\Im}^{a_n,v}c_{\Im}^{\left( a_n-1 \right) N_T+i_{\Im},k} \big] h_{\left( \bar{u}_n-1 \right) N_T+a_n,n_r}+v_{\left( n_r-1 \right) M+m,k},\bigtriangleup f_{\bar{u}_n,a_n}^{}=\bigtriangleup f_{}^{m}\\
				v_{\left( n_r-1 \right) M+m,k},\bigtriangleup f_{\bar{u}_n,a_n}^{}\ne \bigtriangleup f_{}^{m}\\
			\end{array} 
	\end{align}
	\hrule
\end{figure*}

Moreover, the channel coefficient $
h_{( \bar{u}_n-1 ) N_T+a_n,n_r}
$ has absorbed the terms of $
e^{-j2\pi ( f_0+\bigtriangleup f_{\bar{u}_n,a_n}^{} ) {\tau _{a_n}}_{,n_R}}
$ here, and for the sake of convenience in the narrative, we still use the symbol $
h_{( \bar{u}_n-1 ) N_T+a_n,n_r}
$. 

\subsubsection{ML detector}
Additionally, the ML detector can be directly utilized to decode the transmitted data in our proposed system at the receiver end. For more detailed information, interested readers are referred to Appendix \ref{refA}. It should be noted that, in order to estimate the unknown parameters, the ML detector requires \(2^p\) searches, which is undoubtedly a significant computational load. Therefore, to make the proposed system practical, a detection algorithm with lower computational complexity must be developed. 
\subsubsection{Despreading-based Low-complexity detector}
The despreading-based low-complexity (DBLC) detector will be divided into three steps. First, the selected frequency offset is estimated by searching for the maximum value in the receiver output. Second, the despreading of the spreading code is performed to achieve demodulation of the spreading code, the selected transmit antenna, and the frequency offset realignment. Finally, the QAM modulation symbols transmitted by each antenna, corresponding to the frequency offset and the spreading code, are estimated using the ML criterion.

Specifically, combining the output signals of the $m$th channel from
all receiving antennas and $K$ chips yields Eq.\eqref{eq18} (shown in the top of next page).
\begin{figure*}
	\begin{equation}
		\label{eq18}
		\tilde{\boldsymbol{y}}_m=\Bigg\{ \begin{array}{c}
			\begin{split}
				\sqrt{\frac{P_S}{N}}\boldsymbol{h}_{\left( \bar{u}_n-1 \right) N,a_n}\left[ x_{\Re}^{a_n,v}\big[ \boldsymbol{c}_{\Re}^{\left( a_n-1 \right) N_T,i_{\Re}} \big] ^T+jx_{\Im}^{a_n,v}\big[ \boldsymbol{c}_{\Im}^{\left( a_n-1 \right) N_T,i_{\Im}} \big] ^T \right] +\boldsymbol{v}_m\in \mathbb{C} ^{N_R\times K},\bigtriangleup f_{\bar{u}_n,a_n}^{}&=\bigtriangleup f_{}^{m}\\
				\boldsymbol{v}_m\in \mathbb{C} ^{N_R\times K},\bigtriangleup f_{\bar{u}_n,a_n}^{}&\ne \bigtriangleup f_{}^{m}\\
			\end{split}
		\end{array} 
	\end{equation}
	\hrule
\end{figure*}

Next, one can easily obtain the following set:
\begin{equation}
\mathscr{Y} =\{ \left\| \tilde{\boldsymbol{y}}_{1}^{} \right\| ^2,\cdots ,\left\| \tilde{\boldsymbol{y}}_{m}^{} \right\| ^2,\cdots ,\left\| \tilde{\boldsymbol{y}}_{M}^{} \right\| ^2 \}. \nonumber
\end{equation}

Then, \eqref{eq20} gives the expression to estimate the frequency offset as
\begin{equation}
	\label{eq20}
	\begin{split}
		&\left\{ \bigtriangleup f_{\hat{u}_1},\cdots ,\bigtriangleup f_{\hat{u}_n},\cdots ,\bigtriangleup f_{\hat{u}_N} \right\} =arg_N\,\,\mathrm{sort}\left\{ \mathscr{Y} \right\}, 
	\end{split}
\end{equation}
where $arg_N$ denotes frequency offset corresponding to the first $N$ elements taken from the set.

Thus, the corresponding set for the valid channels can be expressed as 
\begin{equation}
\tilde{\mathscr{Y}}=\left\{ \tilde{\boldsymbol{y}}_{\hat{u}_1}^{},\cdots ,\tilde{\boldsymbol{y}}_{\hat{u}_n}^{},\cdots ,\tilde{\boldsymbol{y}}_{\hat{u}_N}^{} \right\}, \nonumber
\end{equation}
or equivalently $\tilde{\mathscr{Y}}=\tilde{\mathscr{Y}}^I+j\tilde{\mathscr{Y}}^Q$,
 where $\tilde{\mathscr{Y}}^I$ and $\tilde{\mathscr{Y}}^Q$ stand for the $I$ and $Q$ components for the set $\tilde{\mathscr{Y}}$. That means 
\begin{align}
	\label{eq23}
	\tilde{\mathscr{Y}}^I=&\{ \tilde{\boldsymbol{y}}_{\hat{u}_1}^{I},\cdots ,\tilde{\boldsymbol{y}}_{\hat{u}_n}^{I},\cdots ,\tilde{\boldsymbol{y}}_{\hat{u}_N}^{I} \}, 
	\\
	\label{eq24}	
	\tilde{\mathscr{Y}}^Q=&\{ \tilde{\boldsymbol{y}}_{\hat{u}_1}^{Q},\cdots ,\tilde{\boldsymbol{y}}_{\hat{u}_n}^{Q},\cdots ,\tilde{\boldsymbol{y}}_{\hat{u}_N}^{Q} \}, 
\end{align}  
where
\begin{align}
	\tilde{\boldsymbol{y}}_{\hat{u}_n}^{I}=&\sqrt{\frac{P_S}{N}}\boldsymbol{h}_{\left( \hat{u}_n-1 \right) N,b_n}^{I}x_{\Re}^{b_n,v}\left[ \boldsymbol{c}_{\Re}^{\left( b_n-1 \right) N_T,i_{\Re}} \right] ^T+\boldsymbol{v}_{\hat{u}_n}^{I},\nonumber
	\\
	\tilde{\boldsymbol{y}}_{\hat{u}_n}^{Q}=&\sqrt{\frac{P_S}{N}}\boldsymbol{h}_{\left( \hat{u}_n-1 \right) N,b_n}^{Q}x_{\Im}^{b_n,v}\left[ \boldsymbol{c}_{\Im}^{\left( b_n-1 \right) N_T,i_{\Im}} \right] ^T+\boldsymbol{v}_{\hat{u}_n}^{Q}.\nonumber
\end{align}
Here, the $\boldsymbol{v}_{\hat{u}_n}^{I}$ and $\boldsymbol{v}_{\hat{u}_n}^{Q}$ are the in-phase and quadrature components for the $\hat{u}_n$th channel AWGN noise, and the power in these matrixes is  $
N_0/M
$. Besides, $
\left\{ b_1,\cdots ,b_n,\cdots b_N \right\} \in \mathbb{Z} ^N
$ are selected from activating antenna pool $
\mathbb{A} $, namely $
b_n\in \left\{ a_1,\cdots, a_N \right\} 
$, which means that the signal involved the frequency increment $
\bigtriangleup f_{\hat{u}_n}
$ is transmitted by the $b_n$th antenna.

Subsequently, we proposed Algorithm \ref{antenna_code} for determining antenna indices and spreading code sequences. To better illustrate this algorithm, we will explain it in detail below using the example of decoding the antenna index $b_1$ and the spreading code sequence for the transmitted frequency offset $\bigtriangleup f_{\hat{u}_1}$.

 In this respect, the product of $\tilde{\boldsymbol{y}}_{\hat{u}_1}^{I}$ and $
\boldsymbol{C}_{\Re}
$
 yields 
\begin{equation}
	\label{eq27}
\begin{split}
	\tilde{\boldsymbol{Y}}_{\hat{u}_1,b_1,i_{\Re}}^{I}=&\tilde{\boldsymbol{y}}_{\hat{u}_1}^{I}\boldsymbol{C}_{\Re}\in \mathbb{C} ^{N_R\times LN_T}
	\\
	=&
	\left[ \tilde{\boldsymbol{Y}}_{\hat{u}_1,1}^{I},\cdots ,\tilde{\boldsymbol{Y}}_{\hat{u}_1,p}^{I},\cdots ,\tilde{\boldsymbol{Y}}_{\hat{u}_1,LN_T}^{I} \right], 
\end{split}
\end{equation}
where
\begin{equation}
\tilde{\boldsymbol{Y}}_{\hat{u}_n,p_1}^{I}=E_c\sqrt{\frac{P_S}{N}}\boldsymbol{h}_{\left( \hat{u}_n-1 \right) N,b_n}^{I}x_{\Re}^{b_n,v}+\boldsymbol{v}_{\hat{u}_n}^{}\boldsymbol{c}_{\Re}^{p},
\end{equation}
when $
p_1=\left( b_1-1 \right) L+i_{\Re}
$ with $
E_c=\sum_{k=1}^K{( c_{\Re}^{\left( n-1 \right) L+i_{\Re},k} ) ^2}
$. Further, if $
p_1\ne \left( b_1-1 \right) L+i_{\Re}
$, the expression of $
\tilde{\boldsymbol{Y}}_{\hat{u}_n,p_1}^{I}
$ is $\tilde{\boldsymbol{Y}}_{\hat{u}_n,p_1}^{I}=\boldsymbol{v}_{\hat{u}_n}^{I}\boldsymbol{c}_{\Re}^{p_1}$,
with $\boldsymbol{v}_{\hat{u}_n}^{I}$ being the real parts of  $\boldsymbol{v}_{\hat{u}_n}^{}$.

Similarly, multipling the matrices $\tilde{\boldsymbol{y}}_{\hat{u}_1}^{Q}$ by the code $\boldsymbol{C}_{\mathfrak{I}}$ gives 
\begin{equation}
	\label{eq28}
	\begin{split}
		\tilde{\boldsymbol{Y}}_{\hat{u}_1,b_1,i_{\mathfrak{I}}}^{Q}=&\tilde{\boldsymbol{y}}_{\hat{u}_1}^{Q}\boldsymbol{C}_{\mathfrak{I}}\in \mathbb{C} ^{N_R\times LN_T}
		\\
		=&\big[ \tilde{\boldsymbol{Y}}_{\hat{u}_1,1}^{Q},\cdots ,\tilde{\boldsymbol{Y}}_{\hat{u}_1,q_1}^{Q},\cdots ,\tilde{\boldsymbol{Y}}_{\hat{u}_1,LN_T}^{Q} \big]. 
	\end{split}
\end{equation}

To estimate the spreading code for the $b_1$th transmitting antenna, we first establish the following vector models $
\{ \| \tilde{\boldsymbol{Y}}_{\hat{u}_1,p_1}^{I} \| ^2 \} _{1}^{LN_T}
$ from the matrix $\tilde{\boldsymbol{Y}}_{\hat{u}_1,b_1,i_{\Re}}^{I}$
and $
\{ \| \tilde{\boldsymbol{Y}}_{\hat{u}_1,q_1}^{Q} \| ^2 \} _{1}^{LN_T}
$ from the matrix $
\tilde{\boldsymbol{Y}}_{\hat{u}_1,b_1,i_{\mathfrak{I}}}^{Q}
$, and then determine the index for maximum value of the elements in these sets  
\begin{align}
	\label{eq29}
	\hat{p}_1=&\underset{p_1}{arg\,\,\max}\{ \| \tilde{\boldsymbol{Y}}_{\hat{u}_1,p}^{I} \| ^2 \}, 
	\\
	\label{eq30}
	\hat{q}_1=&\underset{q_1}{arg\,\,\max}\{ \| \tilde{\boldsymbol{Y}}_{\hat{u}_1,q}^{Q} \| ^2 \}. 
\end{align}

Notice that \eqref{eq29}-\eqref{eq30} have applied the orthogonal condition for Walsh Hadamard codes, namely 
\begin{align}
	( \boldsymbol{c}_{\Re}^{n,i_{\Re}} ) ^T\boldsymbol{c}_{\Re}^{n,i}=\!\bigg\{ \begin{array}{c}
		\!E_c
		,i=i_{\Re}\\
		\!0,i\ne i_{\Re}\\
	\end{array} , \
	( \boldsymbol{c}_{\mathfrak{I}}^{n,i_{\mathfrak{I}}} ) ^T\boldsymbol{c}_{\mathfrak{I}}^{n,i}=\!\bigg\{ \begin{array}{c}
		\!E_c,i=i_{\mathfrak{I}}\\
		\!0,i\ne i_{\mathfrak{I}}\\
	\end{array} . \nonumber
\end{align}

Moreover, one can easily prove that 
\begin{equation}
	\label{eq31}
	\begin{cases}
		\hat{p}_1=\left( b_1-1 \right) L+i_{\Re}\\
		\hat{q}_1=\left( b_1-1 \right) L+i_{\mathfrak{I}}\\
	\end{cases}.
\end{equation}
Hence, the estimates of spreading codes for the $b_1$th transmit antenna can be calculated as 
$\hat{i}_{\Re}^{1}=\mathrm{rem}\left( \hat{p}_1,L \right), 
		\hat{i}_{\mathfrak{I}}^{1}=\mathrm{rem}\left( \hat{q}_1,L\right)$.
Further, we can determine the index of the antenna as 
\begin{equation}
	\label{eq35}
	\hat{b}_1=\mathrm{mod}\left( \hat{p}_1,L \right)+1
\end{equation} 
or 
\begin{equation}
	\label{eq36}
	\hat{b}_1=\mathrm{mod}\left( \hat{q}_1,L \right)+1. 
\end{equation}
Note that either \eqref{eq35} or \eqref{eq36} can be utilized to estimate the antenna index. 

So far, we have demodulated the antenna index of transmitting the frequency offset $
\bigtriangleup f_{\hat{u}_1}^{}
$ and its corresponding spreading code sequence.
However, when either \( \hat{p}_1 \) or \( \hat{q}_1 \) happens to be an integer multiple of \( L \), the aforementioned method concerning encoding and antenna indexing would lead to errors. To rectify this discrepancy, it can be further amended as follows:
\begin{align}
	\begin{cases}
		\hat{i}_{\Re}^{1}=\bigg\{ \begin{array}{r}
			\mathrm{rem}\left( \hat{p}_1,L \right) ,\mathrm{rem}\left( \hat{p}_1,L \right) \ne 0\\
			L,\mathrm{rem}\left( \hat{p}_1,L \right) =0\\
		\end{array} \\
		\hat{i}_{\mathfrak{I}}^{1}=\bigg\{ \begin{array}{r}
			\mathrm{rem}\left( \hat{q}_1,L \right) ,\mathrm{rem}\left( \hat{q},L \right) \ne 0\\
			L,\mathrm{rem}\left( \hat{q}_1,L \right) =0\\
		\end{array} \\
	\end{cases}.
\end{align}
The corresponding estimates for the antenna indices are changed to
\begin{equation}
	\hat{b}_1=\left\{ \begin{array}{r}
		\mathrm{mod}\left( \hat{p}_1,L \right) +1,\mathrm{rem}\left( \hat{p}_1,L \right) \ne 0\\
		\mathrm{mod}\left( \hat{p}_1,L \right) ,\mathrm{rem}\left( \hat{p}_1,L \right) =0\\
	\end{array} \right., 
\end{equation}
or
\begin{equation}
	\label{eq39}
	\hat{b}_1=\left\{ \begin{array}{r}
		\mathrm{mod}\left( \hat{q}_1,L \right) +1,\mathrm{rem}\left( \hat{q}_1,L \right) \ne 0\\
		\mathrm{mod}\left( \hat{q}_1,L \right) ,\mathrm{rem}\left( \hat{q}_1,L \right) =0\\
	\end{array} \right.. 
\end{equation}
Until now, we have estimated the following set of parameters: $
\{ \hat{u}_1,\hat{b}_1,\hat{i}_{\Re}^{1},\hat{i}_{\mathfrak{I}}^{1} \} 
$, but the unknown QAM symbols have not been determined.


\begin{algorithm}
	\caption{Algorithm for determining antenna indices and spreading code sequences}
	\label{antenna_code}
	\begin{algorithmic}[1]
		\STATE Input the valid channel set \eqref{eq23}-\eqref{eq24} and spreading Walsh codes $\boldsymbol{C}_{\Re}$, $
			\boldsymbol{C}_{\mathfrak{I}}$
		\FOR {$n = 1$ to $N$}
		\STATE Build the products $
		\tilde{\boldsymbol{Y}}_{\hat{u}_n,b_n,i_{\Re}}^{I}
		$ and $
		\tilde{\boldsymbol{Y}}_{\hat{u}_n,b_n,i_{\mathfrak{I}}}^{Q}
		$
		like \eqref{eq27} and \eqref{eq28}
		\STATE Establish the vector set $
		\{ \| \tilde{\boldsymbol{Y}}_{\hat{u}_n,p}^{I} \| ^2 \} _{1}^{LN_T}
		$ from the matrix $\tilde{\boldsymbol{Y}}_{\hat{u}_n,b_n,i_{\Re}}^{I}$
		and $
		\{ \| \tilde{\boldsymbol{Y}}_{\hat{u}_n,q}^{Q} \| ^2 \} _{1}^{LN_T}
		$ from the matrix $
		\tilde{\boldsymbol{Y}}_{\hat{u}_n,b_n,i_{\mathfrak{I}}}^{Q}
		$
		\STATE Determine the indexs of the maximum elements from the above two vector sets like 
		\begin{align*}
			\hat{p}_n=&\underset{p}{arg\,\,\max}\{ \| \tilde{\boldsymbol{Y}}_{\hat{u}_n,p}^{I} \| ^2 \} 
			\\
			\hat{q}_n=&\underset{q}{arg\,\,\max}\{ \| \tilde{\boldsymbol{Y}}_{\hat{u}_n,q}^{Q} \| ^2 \} 
		\end{align*}
		\STATE Estimate the indexs of spreading codes 
$$
\begin{cases}
	\hat{i}_{\Re}^{n}=\bigg\{ \begin{array}{r}
		\mathrm{rem}\left( \hat{p}_n,L \right) ,\mathrm{rem}\left( \hat{p}_n,L \right) \ne 0\\
		L,\mathrm{rem}\left( \hat{p}_n,L \right) =0\\
	\end{array} \\
	\hat{i}_{\mathfrak{I}}^{n}=\bigg\{ \begin{array}{r}
		\mathrm{rem}\left( \hat{q}_n,L \right) ,\mathrm{rem}\left( \hat{q}_n,L \right) \ne 0\\
		L,\mathrm{rem}\left( \hat{q}_n,L \right) =0\\
	\end{array} \\
\end{cases}
$$

		\STATE Obtain the antenna index
		$$
		\hat{b}_n=\left\{ \begin{array}{r}
			\mathrm{mod}\left( \hat{p}_n,L \right) +1,\mathrm{rem}\left( \hat{p}_n,L \right) \ne 0\\
			\mathrm{mod}\left( \hat{p}_n,L \right) ,\mathrm{rem}\left( \hat{p}_n,L \right) =0\\
		\end{array} \right. 
		$$
		or
				$$
		\hat{b}_n=\left\{ \begin{array}{r}
			\mathrm{mod}\left( \hat{q}_n,L \right) +1,\mathrm{rem}\left( \hat{q}_n,L \right) \ne 0\\
			\mathrm{mod}\left( \hat{q}_n,L \right) ,\mathrm{rem}\left( \hat{q}_n,L \right) =0\\
		\end{array} \right. 
		$$
		\ENDFOR
		\STATE Output the estimate parameters set: $$
		\mathcal{P} =\big< \{ \hat{u}_1,\hat{b}_1,\hat{i}_{\Re}^{1},\hat{i}_{\mathfrak{I}}^{1} \} ,\cdots ,\{ \hat{u}_N,\hat{b}_N,\hat{i}_{\Re}^{N},\hat{i}_{\mathfrak{I}}^{N} \}  \big> 
		$$ 
	\end{algorithmic}
\end{algorithm}

Next, taking the valid vectors from the set $
\{ \tilde{\boldsymbol{Y}}_{\hat{u}_1,b_1,i_{\Re}}^{I},\cdots ,\tilde{\boldsymbol{Y}}_{\hat{u}_n,b_n,i_{\Re}}^{I},\cdots ,\tilde{\boldsymbol{Y}}_{\hat{u}_N,b_N,i_{\Re}}^{I} \} 
$ yields
\begin{equation}
	\bar{\mathcal{Y}}^I=\{ \tilde{\boldsymbol{Y}}_{\hat{u}_1,\hat{p}_1}^{I}\cdots ,\tilde{\boldsymbol{Y}}_{\hat{u}_n,\hat{p}_n}^{I},\cdots ,\tilde{\boldsymbol{Y}}_{\hat{u}_N,\hat{p}_N}^{I} \} \in \mathbb{R} ^{N_R\times N}.
\end{equation}

Similarly, we have 
\begin{equation}
	\bar{\mathcal{Y}}^Q=\{ \tilde{\boldsymbol{Y}}_{\hat{u}_1,\hat{q}_1}^{Q}\cdots ,\tilde{\boldsymbol{Y}}_{\hat{u}_n,\hat{q}_n}^{Q},\cdots ,\tilde{\boldsymbol{Y}}_{\hat{u}_N,\hat{q}_N}^{Q} \} \in \mathbb{R} ^{N_R\times N}.
\end{equation}
Furthermore, in order to estimate the modulated QAM symbols, Algorithm \ref{qam_code} has been proposed.
\begin{algorithm}
	\caption{Algorithm for determining QAM symbols}
	\label{qam_code}
	\begin{algorithmic}[1]
		\STATE Input the sets $\tilde{\mathcal{Y}}^I$ and $\tilde{\mathcal{Y}}^Q$
		\FOR {$n = 1$ to $N$}
		\STATE Estimate the QAM symbols in $\hat{b}_n$th transmit antenna
		\begin{equation*}
			\begin{split}
				\hat{x}^{\hat{b}_n}=&\underset{v}{arg\,\,\max}\bigg\{ \Big\| \tilde{\boldsymbol{Y}}_{\hat{u}_1,\hat{b}_1,\hat{i}_{\Re}^{1},\hat{p}_1}^{I}+j\tilde{\boldsymbol{Y}}_{\hat{u}_1,\hat{b}_1,\hat{i}_{\mathfrak{I}}^{1},\hat{q}_1}^{Q}  
				\\
				&\qquad\qquad\qquad	 -\sqrt{\frac{P_S}{N}}x_{}^{\hat{b}_n,v}\boldsymbol{h}_{\left( \hat{u}_n-1 \right) N,\hat{b}_n} \Big\| ^2 \bigg\} 
			\end{split}
		\end{equation*}
		\ENDFOR
		\STATE Output the estimate parameters set: $$
		\big\{ \hat{x}^{\hat{b}_1},\cdots ,\hat{x}^{\hat{b}_n},\cdots ,\hat{x}^{\hat{b}_N} \big\} 
		$$
	\end{algorithmic}
\end{algorithm}

Combining the outputs from Algorithm \ref{antenna_code}-\ref{qam_code} gives the all estimated parameters set 
$\tilde{\mathcal{P}}=\left\{ \mathfrak{p} _1,\cdots ,\mathfrak{p} _n,\cdots ,\mathfrak{p} _N \right\}$, 
where the subset $\mathfrak{p}_n$ can be expressed as $\mathfrak{p}_n=\{ \hat{u}_n,\hat{b}_n,\hat{i}_{\Re}^{n},\hat{i}_{\mathfrak{I}}^{n},\hat{x}^{\hat{b}_n} \}$.
Finally, by sorting the subsets in the set $\tilde{\mathcal{P}}$ in ascending order of antennas, we can obtain estimates for $
\{ \hat{a}_1,\cdots \hat{a}_N \} 
$ and $
\{ \hat{\bar{u}}_1,\cdots ,\hat{\bar{u}}_N \} 
$.

\section{Performance Analysis of the GCIM-FORMASM system}
\label{sec3}
In this section, we will conduct a performance analysis of the proposed GCIM system in terms of ABEP, energy efficiency, data rate, and system complexity.

\subsection{ABEP for DBLC}
The expression of the ABEP can be expressed as 
\begin{equation}
	\label{equ46}
P=\frac{P_1p_f+P_2p_c+P_3p_s+P_4p_r +P_5p_m}{p_s+p_f+p_r+p_m+p_c},
\end{equation}  
where \(P_1\) and \(P_2\) denote the ABEP of frequency offset and spreading code index bits, respectively. Additionally, \(P_3\) and \(P_4\) represent the ABEP of the antenna and realigned frequency offset index bits, respectively, with \(P_5\) being the ABEP of constellation bits.

First, we present the derivation for ABEP of frequency offset index bits. In this respect, assume there are  $
C_{p_f}^{q}
$ events by erring with $
q\in \left[ 1,\cdots ,p_f \right] 
$. If frequency offsets cannot be correctly demodulated, it is assumed that the remaining frequency offsets in the frequency offset pool are randomly chosen with equal probabilities. Further, the ABEP model for the frequency offset index bits can be established as  
\begin{equation}
	\label{eq45}
	\begin{split}
		P_1=&\frac{P_f}{\left( 2^{p_f}-1 \right) p_f}\sum_{q=1}^{p_f}{qC_{p_f}^{q}}
=\frac{2^{p_f}P_f}{2\left( 2^{p_f}-1 \right)},
	\end{split}
\end{equation}
where $P_f$ representing the probability of frequency offset not being correctly detected at the receiver. 

To derive \(P_f\), let's first derive the probability \(P_e\) that the output channel containing the transmitting signal is smaller than any channel containing only noise. Recalling \eqref{eq18}, the output of the \(m\)th channel containing a signal can be represented as:
\begin{equation}
	\label{eq46}
	\tilde{\boldsymbol{y}}_m=\sqrt{\frac{P_S}{N}}\boldsymbol{h}_{\left( \bar{u}_n-1 \right) N,a_n}\boldsymbol{d}_{a_n}^{T}+\boldsymbol{v}_m,
\end{equation}
with
\begin{equation}
	\boldsymbol{d}_{a_n}^{}=x_{\Re}^{a_n,v}\boldsymbol{c}_{\Re}^{\left( a_n-1 \right) N_T,i_{\Re}}+jx_{\Im}^{a_n,v}\boldsymbol{c}_{\Im}^{\left( a_n-1 \right) N_T,i_{\Im}}\in \mathbb{C} ^K.
\end{equation}
Further, one can easily obtain $
\left| \boldsymbol{d}_{a_n}^{k} \right|^2=\left| x_{}^{a_n,v} \right|^2,\forall k
$. The corresponding output of the $
m^{\prime}
$th channel containing only noise is $
\tilde{\boldsymbol{y}}_{m^{\prime}}=\boldsymbol{v}_{m^{\prime}}\in \mathbb{C} ^{N_R\times K}
$. Note that each element in the noise matrix \(\boldsymbol{v}_m\) or \(\boldsymbol{v}_{m'}\) is independently and identically distributed (i.i.d.) according to a complex Gaussian distribution, i.e., the \(\left( l,k \right)\)th element has a zero mean and a variance of \(\frac{N_0}{M}\). Hence, each column of $
\tilde{\boldsymbol{y}}_m
$ follows the following distribution: 
\begin{equation}
	\tilde{\boldsymbol{y}}_{m}^{k}\sim \mathcal{C} \mathcal{N} \Big( 0,\big( \frac{P_S}{N}\left| \boldsymbol{d}_{a_n}^{k} \right|^2\sigma ^2+\frac{N_0}{M} \big) \boldsymbol{I} \Big), 
\end{equation}
or equivalently
\begin{equation}
	\tilde{\boldsymbol{y}}_{m}^{k}\sim \mathcal{C} \mathcal{N} \Big( 0,\big( \frac{P_S}{N}\left| x_{}^{a_n,v} \right|^2\sigma ^2+\frac{N_0}{M} \big) \boldsymbol{I} \Big), 
\end{equation} 
with $k=1,2,\cdots ,K$. Then, it can be readily seen that $
\left\| \tilde{\boldsymbol{y}}_m \right\| ^2
$ and $
\left\| \tilde{\boldsymbol{y}}_{m^{\prime}} \right\| ^2
$ follow the central chi-squared distribution with \(
2KN_R
\) degrees of freedom (DoFs) but with different variances, namely the corresponding probability density functions (PDFs) can be given by \cite{proakis2008digital} 
\begin{equation}
	f_{\left\| \tilde{\boldsymbol{y}}_m \right\| ^2}\left( x \right) =\frac{1}{2^{KN_R}\varGamma \left( KN_R \right) \sigma _{1}^{KN_R}}x^{KN_R-1}\exp \big( -\frac{x}{2\sigma _{1}^{}} \big), \nonumber
\end{equation}
and 
\begin{equation}
	f_{\left\| \tilde{\boldsymbol{y}}_{m^{\prime}} \right\| ^2}\left( x \right) =\frac{1}{2^{KN_R}\varGamma \left( KN_R \right) \sigma _{2}^{KN_R}}x^{KN_R-1}\exp ( -\frac{x}{2\sigma _{2}^{}} ). \nonumber
\end{equation}
where $
\sigma _{1}^{}=\frac{P_S}{2N}\left| x_{}^{a_n,v} \right|^2\sigma ^2+\frac{N_0}{2M}
$ and $
\sigma _2=\frac{N_0}{2M}
$. $
\varGamma \left( u \right) =\int_0^{\infty}{t^{u-1}\exp \left( -t \right) dt}
$ denotes the Gamma function.

Define $y=\left\| \tilde{\boldsymbol{y}}_{m^{\prime}} \right\| ^2-\left\| \tilde{\boldsymbol{y}}_m \right\| ^2$.
Further, applying the results [Eq.(4-7)] \cite{simon2002probability} gives the $y$'s PDF  
\begin{align}
		f( y| x_{}^{a_n,v} ) =&\frac{1}{2\sigma _2}\exp \big( -\frac{y}{2\sigma _2} \big) \frac{1}{\varGamma ( KN_R )}\big( \frac{\sigma _2}{\sigma _1+\sigma _2} \big) ^{KN_R}\nonumber
		\\
		&\times\sum_{i=0}^{KN_R-1}{\frac{\varGamma \left( 2KN_R-i-1 \right)}{\varGamma \left( i+1 \right) \varGamma \left( KN_R-i \right)}}\nonumber\\
  &\times\Big( \frac{\sigma _1}{\sigma _1+\sigma _2} \Big) ^{KN_R-i-1}\big( \frac{y}{2\sigma _2} \big) ^i,y>0.
\end{align}


Moreover, using the results [Eq.(3-35)] \cite{GradshteynRyzhik2007TableofIntegrals} yields the cumulative distribution function (CDF) conditioned on $x_{}^{a_n,v}$, given by
\begin{equation}
	\label{eq54}
	\begin{split}
		F\left( y>0\left| x_{}^{a_n,v} \right. \right) =&\int_0^{\infty}{f\left( y\left| x_{}^{a_n,v} \right. \right)}dy,
	\end{split}
\end{equation}
where $
B\left( u,v \right) =\frac{\varGamma \left( u \right) \varGamma \left( v \right)}{\varGamma \left( u+v \right)}
$ stands for the Beta function.
 
Finally, by averaging \eqref{eq54} over $x_{}^{a_n,v}$ results in the expression of \(P_e\), namely
\begin{align}
		P_e=&\frac{1}{J}\sum_{j=1}^J{F( y>0| x_{}^{a_n,v}=x_{}^{a_n,j} )}\nonumber
		\\
		=&\frac{1}{J}\sum_{j=1}^J{\bigg\{ \Big( \frac{\sigma _2}{\sigma _1+\sigma _2} \Big) ^{KN_R} }\sum_{i=0}^{KN_R-1}{\Big( \frac{\sigma _1}{\sigma _1+\sigma _2} \Big) ^{KN_R-i-1}}\nonumber
		\\
		&\times \frac{1}{B( KN_R,KN_R-i ) ( 2KN_R-i )} \bigg\}, 
\end{align} 
with $
x_{}^{a_n,j}
$ denotes the $j$th QAM symbol transmitted by $a_n$th antenna.

Furthermore, an upper bound for the misdetection probability of the transmission frequency increment can be obtained as follows:
\begin{equation}
	\label{eq56}
	P_f\leqslant 1-\left[ \left( 1-P_e \right) ^{M-N} \right] ^N .
\end{equation}
Inserting \eqref{eq56} into \eqref{eq45} gives the ABEP for frequency offset index bits. 

Next, we can focus our attention on deriving the ABEP  for spreading code index bits \( P_2 \). Then, \( P_2 \) can be expressed as \cite{cai2019multi}
\begin{equation}
	\label{eq59}
P_2=\frac{LN_T-1}{LN_T}P_{f1}+\left( 1-P_{f1} \right) \frac{P_C}{\log _2L},
\end{equation}
%
%
where $
P_{f1}=1-\left( 1-P_e \right) ^{M-N}
$ representing the probability of a certain frequency offset being incorrectly detected. The first term on the right side of \eqref{eq59} shows that even in the case of a frequency offset estimation error, there is still a probability of $\frac{1}{LN_T}$ of correctly estimating the Walsh code. $P_c$ is the probability of estimating the corresponding Walsh code incorrectly, even when the frequency offset is correctly estimated in the first step.  

Since the $I$ and $Q$ channels are decoded independently, it follows that: $
P_c=\frac{1}{2}P_{c}^{I}+\frac{1}{2}P_{c}^{Q}
$ where $P_{c}^{I}$ and $P_{c}^{Q}$ denote the represent the average probability of erroneous detection of spreading code indices for the $I$ and $Q$ components, respectively.

To derive \( P_{c}^{I} \), we can first derive the probability $P_{e}^{w}$ that a column in matrix $\tilde{\boldsymbol{Y}}_{\hat{u}_n,b_n,i_{\Re}}^{I}$ containing only noise is smaller than a column from $\tilde{\boldsymbol{Y}}_{\hat{u}_n,b_n,i_{\Re}}^{I}$ containing  signal plus noise.
Recall the Algorithm \ref{antenna_code}, one can easily obtain the $p$th column output including the signal from the matrix $\tilde{\boldsymbol{Y}}_{\hat{u}_n,b_n,i_{\Re}}^{I}$, as 
\begin{equation}
	\tilde{\boldsymbol{Y}}_{\hat{u}_n,p}^{I}=E_c\sqrt{\frac{P_S}{N}}\boldsymbol{h}_{\left( \hat{u}_n-1 \right) N,b_n}^{}x_{\Re}^{b_n,v}+\mathbf{v}^p,
\end{equation}
where $
\mathbf{v}^p=\boldsymbol{v}_{\hat{u}_n}^{I}\boldsymbol{c}_{\Re}^{p}\in \mathbb{C} ^{N_R}\sim \mathcal{N} \left( 0,\frac{E_cN_0}{2}\boldsymbol{I} \right) 
$. Similarly, $
p^{\prime}
$th column that contains only noise in matrix $\tilde{\boldsymbol{Y}}_{\hat{u}_n,b_n,i_{\Re}}^{I}$ is $\tilde{\boldsymbol{Y}}_{\hat{u}_n,p^{\prime}}^{I}=\mathbf{v}^{p^{\prime}}$
with $
\mathbf{v}^{p^{\prime}}=\boldsymbol{v}_{\hat{u}_n}^{I}\boldsymbol{c}_{\Re}^{p^{\prime}}\sim \mathcal{N} \left( 0,\frac{E_cN_0}{2}\boldsymbol{I} \right) 
$. Subsequently, we have 
\begin{equation}
\tilde{\boldsymbol{Y}}_{\hat{u}_n,p}^{I}\sim \mathcal{N} \Big( E_c\sqrt{\frac{P_S}{N}}\boldsymbol{h}_{\left( \hat{u}_n-1 \right) N,b_n}^{}x_{\Re}^{b_n,v},\frac{E_cN_0}{2}\boldsymbol{I} \Big). 
\end{equation}
Thus, $
\| \tilde{\boldsymbol{Y}}_{\hat{u}_n,p}^{I} \| ^2
$ and $
\| \tilde{\boldsymbol{Y}}_{\hat{u}_n,p^{\prime}}^{I} \| ^2
$ follow an non-central chi-squared distribution with $N_R$ DoFs and a central chi-squared distribution with $N_R$ DoFs, respectively. Further, their corresponding PDF expressions are \cite{proakis2008digital}
\begin{align}
	f_{\left\| \tilde{\boldsymbol{Y}}_{\hat{u}_n,p}^{I} \right\| ^2}\left( p \right) =&\frac{1}{2\sigma _{3}^{2}}\big( \frac{p}{s^2} \big) ^{\frac{N_R-2}{4}}\exp \big( -\frac{p+s^2}{2\sigma _{3}^{2}} \big)\nonumber\\
 &\qquad\qquad\quad\qquad\times I_{\frac{N_R}{2}-1}\big( \frac{s}{\sigma _{3}^{2}}\sqrt{p} \big),
\end{align}
and 
\begin{equation}
	f_{\left\| \tilde{\boldsymbol{Y}}_{\hat{u}_n,p^{\prime}}^{I} \right\| ^2}\left( x \right) =\frac{x^{N_R/2-1}}{2^{N_R/2}\varGamma \left( N_R/2 \right) \sigma _{3}^{N_R}}\exp \big( -\frac{x}{2\sigma _{3}^{2}} \big), 
\end{equation}
with $
s=\sqrt{\sum_{k=1}^{N_R}{\big( E_{c}^{}\sqrt{\frac{P_S}{N}}x_{\Re}^{b_n,v}h_{\left( \hat{u}_n-1 \right) N+b_n,k} \big) ^2}}
$ [2.3-30] \cite{proakis2008digital} and 
$
\sigma _{3}^{2}=\frac{E_cN_0}{2}
$. Moreover, $
I_{\alpha}\left( p \right) 
$ denotes the modified Bessel function of the first kind with the order $\alpha$.

Further, the conditional probability based on \(s\) and \(x_{\Re}^{b_n,v}\) can be expressed as:
\begin{equation}
	P_{e}^{w}( s,x_{\Re}^{b_n,v} ) =Pr\big( \| \tilde{\boldsymbol{Y}}_{\hat{u}_n,p^{\prime}}^{I} \| ^2<\| \tilde{\boldsymbol{Y}}_{\hat{u}_n,p}^{I} \| ^2,\forall p^{\prime}\ne p\big| s,x_{\Re}^{b_n,v}  \big) ,\nonumber
\end{equation}
where $ {Pr}(\bullet) $denotes the probability of correct symbol detection. Since the events represented by \( {Pr}(\bullet) \) are dependent due to the presence of $
\| \tilde{\boldsymbol{Y}}_{\hat{u}_n,p^{\prime}}^{I} \| ^2
$, one can introduce a condition on $
\| \tilde{\boldsymbol{Y}}_{\hat{u}_n,p^{\prime}}^{I} \| ^2
$ to render these events independent. That is
\begin{equation}
	\begin{split}
&P_{e}^{w}( s,x_{\Re}^{b_n,v} ) =\int_0^{\infty}{\Big[ Pr\big( \| \tilde{\boldsymbol{Y}}_{\hat{u}_n,p^{\prime}}^{I} \| ^2< }\nonumber
\\
&\qquad\quad\| \tilde{\boldsymbol{Y}}_{\hat{u}_n,p}^{I} \| ^2| \| \tilde{\boldsymbol{Y}}_{\hat{u}_n,p}^{I} \| ^2,s,x_{\Re}^{b_n,v}  \big) \Big] ^{LN_T-1}f_{\| \tilde{\boldsymbol{Y}}_{\hat{u}_n,p}^{I} \| ^2}( p ) dp.
	\end{split}
\end{equation}
Hence, we have 
\begin{align}
	\label{equ68}
	&\quad P_{e}^{w}( s,x_{\Re}^{b_n,v} ) \nonumber\\
 & =\int_0^{\infty}{\Big[ \int_0^p{f_{\| \tilde{\boldsymbol{Y}}_{\hat{u}_n,p^{\prime}}^{I} \| ^2}( x )} \Big] ^{LN_T-1}}f_{\| \tilde{\boldsymbol{Y}}_{\hat{u}_n,p}^{I} \| ^2}( p ) dp.
\end{align}

Inserting the PDFs of $
\| \tilde{\boldsymbol{Y}}_{\hat{u}_n,p}^{I} \| ^2
$ and $
\| \tilde{\boldsymbol{Y}}_{\hat{u}_n,p^{\prime}}^{I} \| ^2
$ into \eqref{equ68} yields the expression of condition probability $P_{e}^{w}( s,x_{\Re}^{b_n,v} )$.

Next, we can shift our focus to \(s\). In this respect, it is easy to see that \(s\) follows a generalized Rayleigh distribution [2.3-51] \cite{proakis2008digital}, with its PDF expressed as:  
\begin{equation}
	\label{equ69}
	f\left( s \right) =\frac{s^{N_R-1}}{2^{\frac{N_R-2}{2}}\sigma _{s}^{N_R}\varGamma \left( \frac{N_R}{2} \right)}e^{-\frac{s^2}{2\sigma _{s}^{2}}},s\geqslant 0,
\end{equation}
with $
\sigma _{s}^{2}=\frac{E_{c}^{2}P_S}{2N}| x_{\Re}^{b_n,v}|^2\sigma 
$. Combining \eqref{equ68} and \eqref{equ69} results in the conditional probability  \(x_{\Re}^{b_n,v}\), namely
\begin{equation}
	\label{eq70}
	P_{e}^{w}( x_{\Re}^{b_n,v} ) =\int_0^{\infty}{P_{e}^{w}( s,x_{\Re}^{b_n,v} ) f( s )}ds.
\end{equation}

By averaging \eqref{eq70} over $x_{\Re}^{b_n,v}$ gives  \(P_e^w\), as
\begin{equation}
P_{e}^{w}=\frac{1}{J}\sum_{j=1}^J{P_{e}^{w}\left( x_{\Re}^{b_n,v}\left| x_{\Re}^{b_n,v}=x_{\Re}^{b_n,j} \right. \right)}.
\end{equation}

One can easily prove that $P_{c}^{I}=1- P_{e}^{w} $.

Next, we will derive the expression for the probability \( P_3 \). By observing Algorithm 1, we can conclude that the correct demodulation of the antenna depends on whether the Walsh code of either the $I$ or $Q$ channel is demodulated correctly. Thus, \( P_3 \) can be expressed as 
\begin{equation}
	\label{eq73}
	P_3=1-\left( 1-P_w \right) ^N,
\end{equation}
with $P_w=\frac{LN_T-1}{LN_T}P_{f1}+\left( 1-P_{f1} \right) {P_C}$.

Subsequently, the following part will provide the derivation of  \( P_4 \). By reviewing the corresponding parts in Algorithm 1, we can observe that the prerequisite for correct frequency offset rearrangement demodulation is the correct demodulation of both the frequency offset and the indexes of antennas\footnote{This paper does not consider the following scenario: the frequency offset is correctly demodulated, but the chosen antenna is not correctly demodulated even though its index order is correct. For example, if the transmitted frequency offsets are \(\{\bigtriangleup f_1, \bigtriangleup f_2, \bigtriangleup f_3\}\) and the chosen antenna indices are \(\{3, 1, 5\}\), the rearranged frequency offset pool would be \(\{\bigtriangleup f_2, \bigtriangleup f_1, \bigtriangleup f_3\}\). Additionally, on the receiving end, if the demodulated frequency offsets are \(\{\bigtriangleup f_1, \bigtriangleup f_2, \bigtriangleup f_3\}\) and the antenna indices are \(\{4, 2, 6\}\), the further rearranged demodulated frequency offsets would be \(\{\bigtriangleup f_2, \bigtriangleup f_1, \bigtriangleup f_3\}\).}. Hence, the expression for \( P_4 \) can be given by
\begin{align}
	\label{equ75}
	P_4=1-\left[ \left( 1-P_{f1} \right) \left( 1-P_C \right) \right] ^N.
\end{align}

Finally, we can focus on determining the probability \( P_5 \). The correct demodulation of QAM symbols first requires ensuring the antenna is correctly demodulated. Hence, the ABEP for QAM symbols can be calculated as 
\begin{align}
	\label{eq76}
	P_5=\frac{1}{2}P_w+\left( 1-P_w \right) P_{QAM},
\end{align}
where $P_{QAM}$ represents the ABEP of constellation bits when the frequency offset index, Walsh code index, and antenna index are all accurately detected.

Consider a rectangular QAM signal, which can be represented as an $I$ channel signal modulated by \( \alpha \)-pulse amplitude modulation (PAM) and a $Q$ channel signal modulated by \( \beta \)-PAM, namely $
J=\alpha \times \beta 
$. Then, the error probability for the \( l \)th bit of the phase component can be expressed as \cite{cho2002general}:
\begin{align}
	\label{eq75}
		P_I( l| \gamma ) =&\frac{2}{\alpha}\sum_{i=0}^{( 1-2^{-l} ) \alpha -1}{\bigg\{ ( -1 ) ^{\lfloor \frac{i\cdot 2^{l-1}}{\alpha} \rfloor}\Big[ 2^{l-1}-\lfloor \frac{i\cdot 2^{l-1}}{\alpha}+\frac{1}{2} \rfloor \Big] }\nonumber
		\\
		\,\,            &
 \times Q\Big( \left( 2i+1 \right) \sqrt{\frac{6\log _2\left( \alpha \beta \right) \gamma}{\alpha ^2+\beta ^2-2}} \Big) \bigg\}, 
\end{align}
with $Q\left( x\right) $ being the $Q$ function and $	\gamma =\frac{ME_{c}^{}P_S| x_{}^{\hat{b}_n,v} |^2}{NN_0}\| \boldsymbol{h}_{( \hat{u}_n-1 ) N,\hat{b}_n} \| ^2$.
Moreover, $\gamma$ obeys a central chi-square distribution with $2N_R$ DoFs, and its PDF can be written as 
\begin{equation}
	\label{eq77}
	f\left( \gamma \right) =\frac{1}{2^{N_R}\varGamma \left( N_R \right) \sigma _{5}^{N_R}}\gamma ^{N_R-1}\exp ( -\frac{\gamma}{2\sigma _{5}^{}} ), 
\end{equation} 
with $
\sigma _{5}^{}=\frac{ME_{c}^{}P_S\left| x_{}^{\hat{b}_n,v} \right|^2\sigma ^2}{2NN_0}
$.
Applying \eqref{eq75} and \eqref{eq77} yields the CDF conditioned on $x_{}^{\hat{b}_n,v}$, i.e.,
\begin{align}
	\label{eq78}
		&P_I( l| x_{}^{\hat{b}_n,v} ) =\int_0^{\infty}{P_I( l| \gamma ) f_{\gamma}( \gamma )}d\gamma \nonumber
		\\
		&\qquad=\frac{2}{\alpha}\sum_{i=0}^{( 1-2^{-l} ) \alpha -1}{\Big\{ ( -1 ) ^{\lfloor \frac{i\cdot 2^{l-1}}{\alpha} \rfloor}\Big[ 2^{l-1}-\lfloor \frac{i\cdot 2^{l-1}}{\alpha}+\frac{1}{2} \rfloor \Big] }\nonumber
		\\
		&\qquad
		 \times \left[ P\left( c \right) \right] ^{N_R}\sum_{k=0}^{N_R-1}{C_{N_R-1+k}^{k}}\left[ 1-P\left( c \right) \right] ^k \Big\}, 
\end{align}
with $
P\left( x \right) =\frac{1}{2}( 1-\sqrt{\frac{x}{1+x}} ) ,x>0
$ with $c=
\left( 2i+1 \right) ^2\frac{3\log _2\left( \alpha \beta \right) \sigma _{5}^{}}{\alpha ^2+\beta ^2-2}
$. Note that we utilized the results in Appendix B of Reference \cite{alouini1999unified} in deriving the above equation.

By averaging \eqref{eq78} on $
x_{}^{\hat{b}_n,v}
$ gives  $		P_I( l ) =\frac{1}{J}\sum_{j=1}^J{P_I( l| x_{}^{\hat{b}_n,v}=x_{}^{\hat{b}_n,j} )}.$
Similarly, we can obtain the error rate expression for $l$th bits in the quadrature components as follows:
\begin{align}
&
P_Q\left( l \right) =\frac{1}{J}\sum_{j=1}^J{\frac{2}{\beta}\sum_{i=0}^{( 1-2^{-l} ) \beta -1}{\bigg\{ ( -1 ) ^{\lfloor \frac{i\cdot 2^{k-1}}{\beta} \rfloor} }\Big[ 2^{k-1}- }\nonumber
\\
& \lfloor \frac{i\cdot 2^{k-1}}{\beta}+\frac{1}{2} \rfloor \Big] [ P( c ) ] ^{N_R}\sum_{k=0}^{N_R-1}{C_{N_R-1+k}^{k}}[ 1-P( c) ] ^k  \bigg\} .
\end{align}
Subsequently, the expression for $
P_{QAM}
$ can be derived as follows:
\begin{equation}
	P_{QAM}=\frac{1}{\log _2( \alpha \beta )}\Big( \sum_{l=1}^{\log _2( \alpha )}{P_I( l ) +\sum_{l=1}^{\log _2( \beta )}{P_Q( l )}} \Big). 
\end{equation}

Applying \eqref{eq45}, \eqref{eq59}, \eqref{eq73}, \eqref{equ75},  \eqref{eq76} and \eqref{equ46} gives the final expression of ABEP for the proposed system.

\subsection{Energy Efficiency and Data Rate}
In the proposed GCIM-FORMASM system, \( p_m \) bits of data are directly transmitted using QAM modulation, while \( p_s \) bits of data are used to select the transmitting antenna index. Additionally, \( p_f \) and \( p_r \) bits of data represent the selected frequency offset and frequency offset rearrangement, respectively, and the remaining \( p_c \) bits form the spreading code sequence. Therefore, when transmitting \( p \) bits of data, only \( p_m \) bits require energy for transmission, while the remaining bits \( p_s+p_f+p_r+p_c \) do not require energy. The energy savings ratio for transmitting \( p \) bits of data can thus be expressed as $E_{sav}=( 1-\frac{p_m}{p} ) E_b\%$
with the bit energy is denoted by $E_b$, whose expression is $
E_b=\frac{E_xE_c}{p}
$ with $E_x$ being the average energy of $J$-ary QAM modulation symbol. 
\begin{table}[htp]
		\caption{Energy Saving Comparisons of GCIM-FORMASM Systems Compared to FOPIM, GSCIM GCIM-SM and SM}
		\vspace{0pt}
		\centering
  \resizebox{0.49\textwidth}{12mm}{
	\begin{tabular}{|c|c|c|c|c|c|c|c|c|}
		\hline
		\multicolumn{5}{|c|}{Parameters} & \multicolumn{4}{c|}{Systems} \\
		\hline
		$N_T$ & $N$ & $M$ & $L$& $J$  & FOPIM & GSCIM & \makecell[c]{GCIM\\-SM}&SM\\
		\hline
		4 & 2 & 8 & 8& 8 &	36.00$\%$ & 36.00$\%$ & 44.00$\%$ & 68.00$\%$  \\
		\hline
		8 & 2 & 8 & 8& 4 &	24.00$\%$ & 36.00$\%$ & 48.00$\%$ & 72.00$\%$  \\
		\hline
		4 & 2 & 12 & 8& 4 &	36.00$\%$ & 44.00$\%$ & 52.00$\%$ & 76.00$\%$  \\
		\hline
		6 & 3 & 6 & 4& 2 &	52.00$\%$ & 56.00$\%$ & 64.00$\%$ & 80.00$\%$  \\
		\hline
	\end{tabular}
 }
\label{table1}
\end{table}

Table \ref{table1} presents the energy savings comparison between the proposed GCIM-FORMASM system and the existing FOPIM \cite{jian2023fda}, GSCIM \cite{zhang2021generalized}, GCIM-SM \cite{cogen2020generalized}, and SM \cite{gandotra2017survey} systems. It should be noted that the results shown in Table \ref{table1} are based on the transmission of the same amount of data but with different system configurations. From the table, it is evident that the proposed GCIM-FORMASM system is more energy-efficient compared to the existing modulation schemes, aligning better with the concept of green communications. Particularly, when the system configuration is \(N_T = 6\), \(N = 3\), \(M = 6\), \(L = 4\), and \(J = 2\), the energy efficiency of the GCIM-FORMASM structure can improve by more than double compared to the FOPIM, GSCIM, GCIM-SM, and SM systems. 

\begin{table}[htp]
	\caption{The Comparisons for Date Rate Between Different Systems (Bits  Per Same Symbol Duration)}
	\vspace{0pt}
	\centering
 \resizebox{0.49\textwidth}{11.6mm}{
	\begin{tabular}{|c|c|c|c|c|c|c|c|c|c|}
		\hline
		\multicolumn{5}{|c|}{Parameters} & \multicolumn{5}{c|}{Systems} \\
		\hline
		$N_T$ & $N$ & $M$ & $L$& $J$ &\tabincell{c}{Our\\method} & FOPIM & GSCIM &\tabincell{c}{GCIM\\-SM}  & SM\\
		\hline
		4 & 2 & 8 & 8& 8 &	25 & 22 & 16 & 11 &5 \\
		\hline
		6 & 3 & 6 & 16& 8 &	43 & 27 & 31 & 13 &5 \\
		\hline
		8 & 4 & 8 & 16& 4 &	56 & 31 & 34 & 13 &5 \\
		\hline
	5 & 2 & 12 & 4& 4 &	22 & 25 & 11 & 8 &4\\
		\hline
	\end{tabular} }
	\label{table2}
\end{table}

Further, Table \ref{table2} give the  comparisons for date rate between different systems. From the results in Table \ref{table2}, it can be observed that in most cases, the data rate of the system proposed in this paper is significantly higher compared to existing systems, mainly due to the utilization of frequency offset index, spreading code index, and multi-antenna transmission. This is especially noticeable when the number of antennas and spreading codes is large. However, from the last column of the table, it is noted that when the number of spreading codes is small, the data rate of the system proposed in this paper may be lower than FOPIM, but higher than other systems. This is expected, as FOPIM activates all antennas to transmit data, while this paper only activates a subset of antennas.
\subsection{System Complexity Analysis}
This subsection will analyze the algorithmic complexity of the proposed ML and DBLC detectors. Specifically, reviewing \eqref{eq99} of the ML detector, it is evident that each search of the ML detector requires $
N_TMN_RK+2NN_TMK+N_RK
$ multiplications, where $
N_TMN_RK+2NN_TMK$ denotes the complexity of calculating the term  $
\boldsymbol{H}\left( \boldsymbol{G}^I\boldsymbol{X}^I+j\boldsymbol{G}^Q\boldsymbol{X}^Q \right) 
$ as well as $N_RK$ is the complexity of performing $
\left\| \cdot \right\| ^2
$ operations. Therefore, implementing the ML detector requires a total complexity of being 
\begin{equation}
	\label{eq85}
	\mathcal{O} \left[ \left( N_TMN_RK+2NN_TMK+N_RK \right) 2^p \right] 
\end{equation}

The DBLC detector requires \( 2KMN_R \) multiplications to demodulate the transmitted frequency offset. Further, Algorithm \ref{antenna_code} indicates that calculating $
\{ \| \tilde{\boldsymbol{Y}}_{\hat{u}_n,p}^{I} \| ^2 \} _{1}^{LN_T}
$  requires $
( N_R+1 ) KLN_T
$ multiplications. Therefore, the computational complexity for determining the sequences of spreading code and antenna  is: $
\mathcal{O} [ 2\left( N_R+1 \right) KLN_TN ] 
$. Subsequently, Algorithm \ref{qam_code} implies that the multiplications of computing the term 
$
\big\| \tilde{\boldsymbol{Y}}_{\hat{u}_1,\hat{b}_1,\hat{i}_{\Re}^{1},\hat{p}_1}^{I}+j\tilde{\boldsymbol{Y}}_{\hat{u}_1,\hat{b}_1,\hat{i}_{\mathfrak{I}}^{1},\hat{q}_1}^{Q}  -\sqrt{\frac{P_S}{N}}x_{}^{\hat{b}_n,v}\boldsymbol{h}_{\left( \hat{u}_n-1 \right) N,\hat{b}_n} \big\| ^2
$ are $N_R$. Hence, the computational complexity required for detecting the transmitted QAM symbols is: $
\mathcal{O} \left[ NJN_R \right] 
$. Therefore, the complexity required for the DBLC detector can be expressed as:
\begin{equation}
	\label{eq86}
	\mathcal{O} \left[ 2KMN_R+\left( N_R+1 \right) KLN_TN+NJN_R \right] 
\end{equation}
Comparing \eqref{eq85} and \eqref{eq86}, it is evident that the complexity of the proposed DBLC algorithm is significantly lower than that of the ML method. Therefore, the DBLC algorithm is more practical for real-world applications.
\section{Simulation Results}
\label{sec4}
To validate the effectiveness of our proposed approaches, we present extensive numerical simulation results in this section. Unless otherwise specified, the simulation parameters are designed as follows: $
f_0=3\mathrm{GHz}
$, $P_s=1$, $
\bigtriangleup f=20\mathrm{KHz}
$\footnote{For convenience, let's assume a linear frequency offset between array elements.}, $
\left( R,\theta \right) =\left( 300\mathrm{m},60^\circ \right) 
$ and $\sigma ^2=1$. In the figure, 'Ana' and 'Sim' represent the results obtained through theoretical analysis and Monte Carlo (MC) simulations, respectively. Besides, the definition of signal-to-noise ratio (SNR) in decibels (dB) is given by $\mathrm{SNR}=10\log \frac{P_s^2}{N_0}$.

Additionally, to highlight the superiority of the system proposed in this paper, other existing systems such as  FOPIM \cite{jian2023fda}, GSCIM \cite{zhang2021generalized}, GCIM-SM \cite{cogen2020generalized}, and SM \cite{gandotra2017survey} will be used for comparison. It should be noted that if all the antennas in the proposed system transmit the same carrier frequency, the GCIM-FORMASM system degenerates into the existing GSCIM system. However, we find that the algorithm used in GSCIM proposed in \cite{zhang2021generalized} for detecting spreading code sequences and the number of transmitting antennas is relatively complex. Therefore, applying the proposed Algorithms 1 and 2 to GSCIM can significantly reduce computational complexity.
To distinguish it from the GSCIM system, we name it the generalized code index modulation-aided multiple-antenna spatial modulation  (GCIM-MASM) system in this paper.

\begin{figure}[htp]
	\centering
	\subfigure[]{
		{\includegraphics[width=0.3\textwidth]{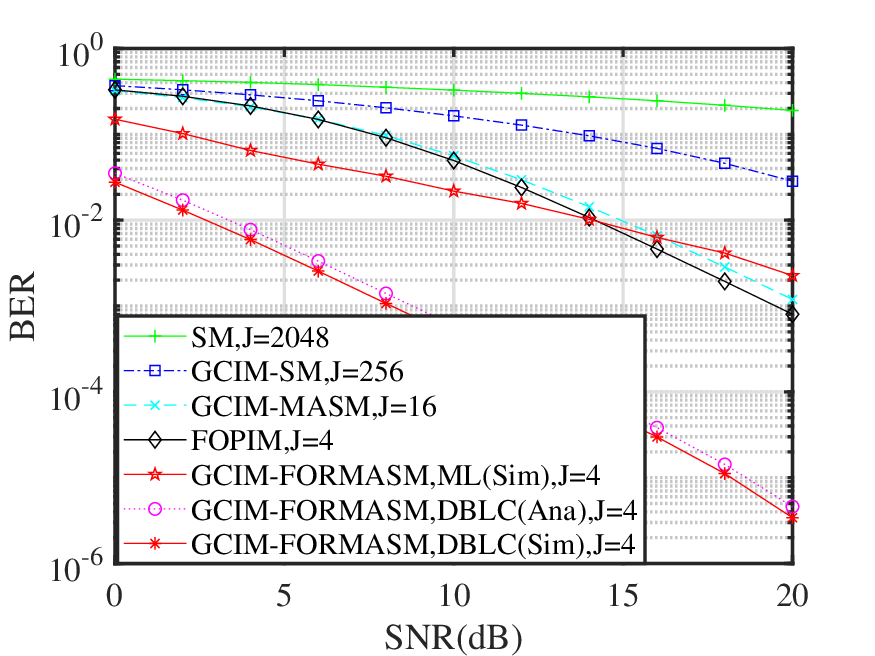}}}
	\subfigure[]{
		{\includegraphics[width=0.3\textwidth]{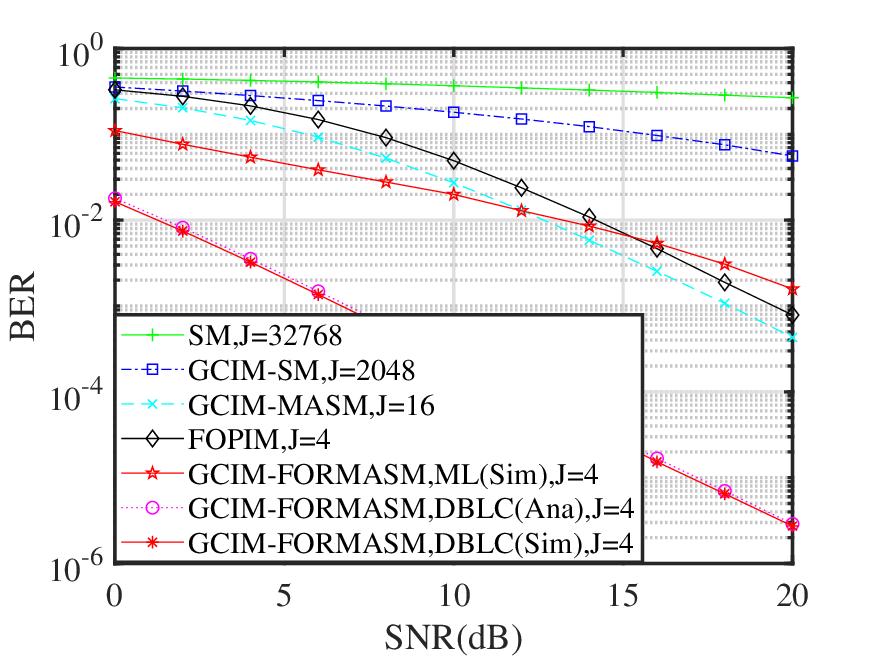}}}
	\caption{BER performance for considered systems versus SNR when $N_T=4$,$N=2$, $M=4$ $N_R=2$. (a) $L=2$. (b) $L=4$.}
	\label{fig44}
\end{figure}
The BER performance versus SNR curves under the assumption of $N_T=4$, $N_R=2$ are presented, as shown in Fig. \ref{fig44}. To ensure a fair comparison, let's assume that all the systems considered in this paper transmit the same number of data bits. Consequently, the QAM modulation orders for the SM, GCIM-SM, GCIM-MASM, and FOPIM systems are 2048, 256, 16, and 4, respectively, in Fig. \ref{fig44}(a). From the figure, it is evident that under these parameters, the SM system has the worst BER performance, while the FOPIM and GCIM-MASM systems perform similarly. The proposed system consistently outperforms the traditional systems, with the DBLC algorithm showing the best performance. For instance, at an SNR of 20 dB, the BER of the DBLC algorithm is below \(10^{-5}\), whereas the BER of the existing systems is above \(10^{-3}\). In other words, the proposed algorithm achieves a performance improvement of approximately 100 times. Furthermore, comparing the DBLC algorithm with the ML algorithm reveals that the DBLC algorithm outperforms the ML algorithm.

\begin{figure*}[htp]
	\centering
	\subfigure[]{
		{\includegraphics[width=0.322\textwidth]{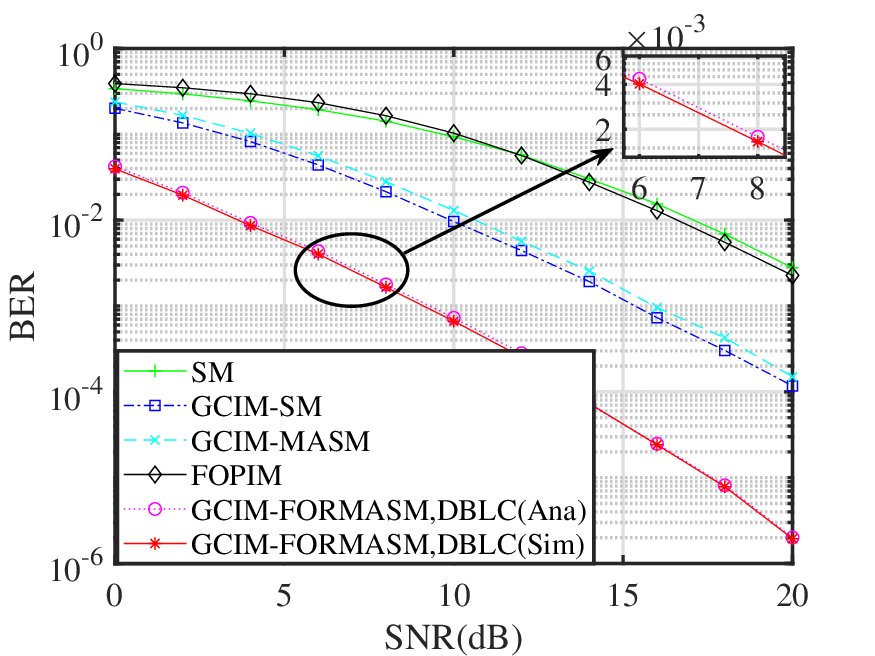}}}
	\subfigure[]{
			{\includegraphics[width=0.322\textwidth]{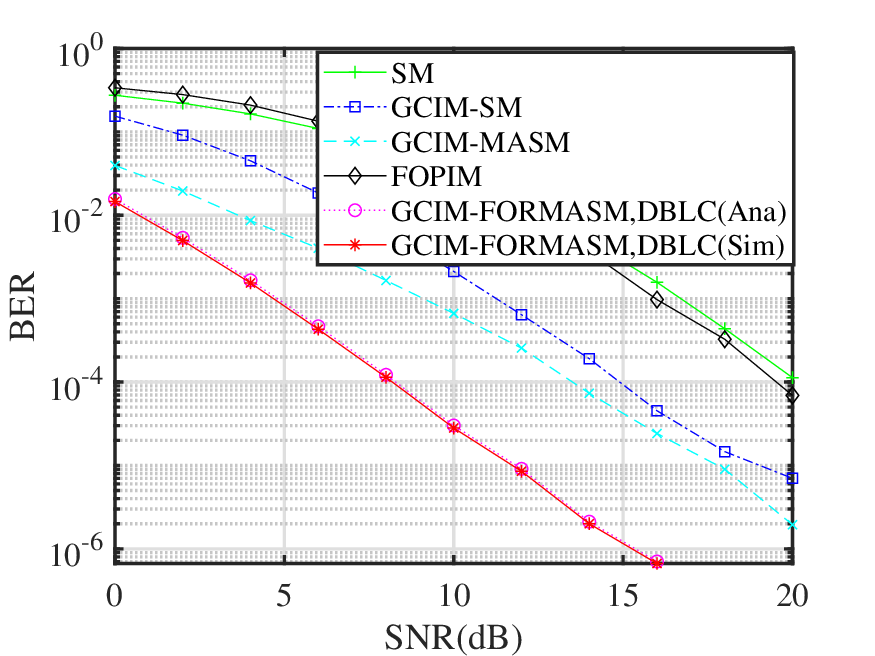}}}
	\subfigure[]{
				{\includegraphics[width=0.322\textwidth]{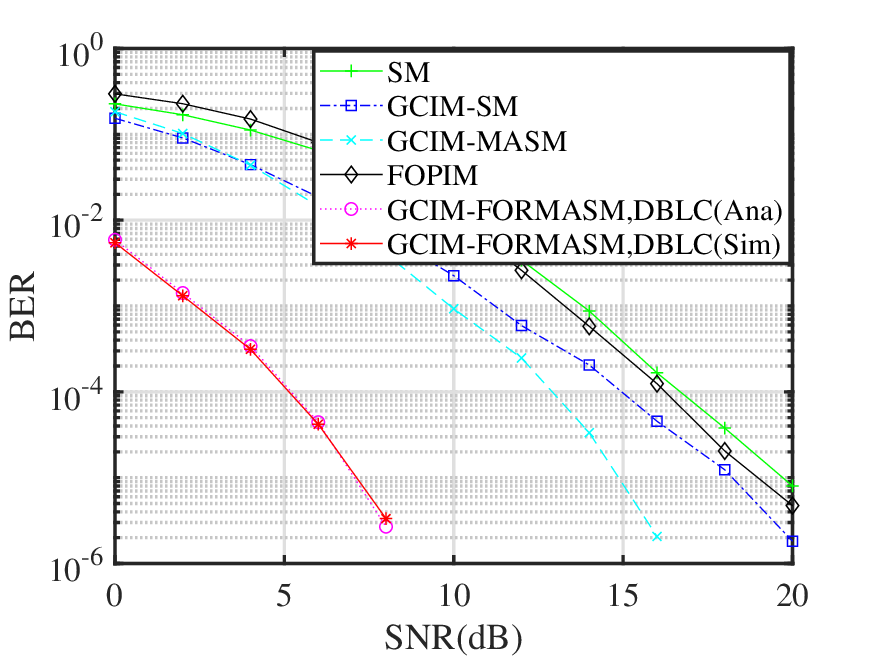}}}
	\caption{Comparison of BER performance for different systems with various receiving antennas, $M=8$, $N_T=4$, $N=3$, $L=8$, $J=8$. (a) $N_R=2$. (b) $N_R=3$. (c) $N_R=4$.}
	\label{fig4}
\end{figure*}

Fig.\ref{fig4} shows the BER performance comparison of different systems with various numbers of receiving antennas. In both the GCIM-FORMASM and GCIM-MASM systems, only three out of five transmitting antennas are activated. From the results, it can be observed that the BER values obtained from the DBLC algorithm's theoretical upper bound and Monte Carlo simulations match well, fully demonstrating the correctness of the theoretical analysis presented earlier. Fig.\ref{fig4}(a) indicates that, under the same system architecture and with the transmission of the same modulation order QAM symbols, the proposed GCIM-FORMASM system has the best BER performance\footnote{This comparison is actually unfair because the GCIM-FORMASM system transmits 36 bits, while the SM  GCIM-SM, GCIM-MASM, and FOPIM  systems transmit only 4, 11, 21 and 22 bits of data, respectively. However, even under such stringent conditions, the performance of the proposed system remains the best.}. The reason for this result is that the proposed system benefits from the Walsh code processing gain compared to the FOPIM system, and it gains from multi-channel processing compared to the GCIM-SM and GCIM-MASM systems. Finally, compared to GCIM-FORMASM, the SM system lacks both the spreading code gain and the multi-channel processing gain at the receiver. Therefore, even under such stringent conditions, the proposed system still exhibits superior BER performance. Meanwhile, the BER performance of FOPIM and SM is almost the same\footnote{The reason this result differs from the conclusion of reference \cite{jian2023fda} is that \cite{jian2023fda} assumes that FOPIM and SM transmit the same amount of data bits.} and is the worst among the systems considered in this paper.

Unlike Fig.\ref{fig4}(a), Fig.\ref{fig4}(b) and Fig.\ref{fig4}(c) present the results for different numbers of receiving antennas. Comparing the three subplots, it can be seen that as the number of receiving antennas increases, the performance of all systems improves. This phenomenon is normal because, in MIMO/SIMO communication systems, an increase in the number of receiving antennas results in greater reception gains. 

\begin{figure}[htp]
	\centering
	\subfigure[]{
		{\includegraphics[width=0.3\textwidth]{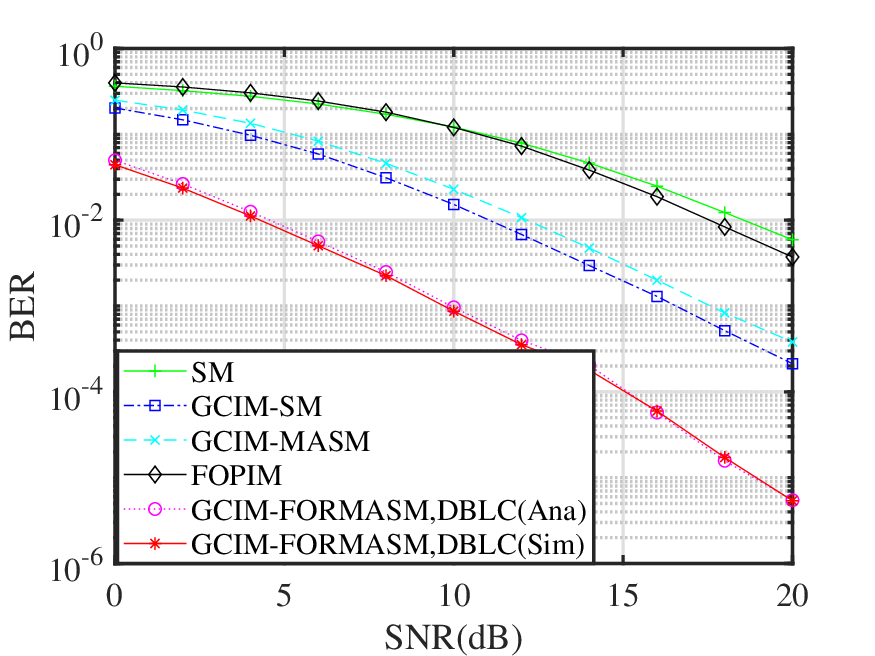}}}
	\subfigure[]{
		{\includegraphics[width=0.3\textwidth]{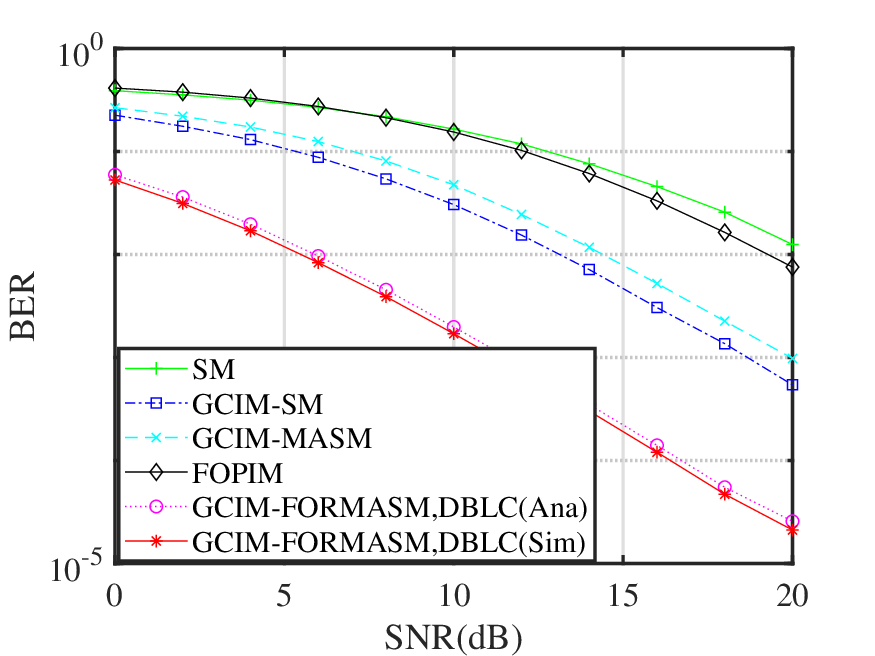}}}
	\caption{Performance comparison under different QAM modulation orders, $M=8$, $N_T=4$, $N=3$, $L=8$, $N_R=2$. (a) $J=16$. (b) $J=32$. }
	\label{fig5}
\end{figure}

Furthermore, Fig.\ref{fig5} presents the BER performance results of different systems under various QAM modulation orders. From the comparison of Fig.\ref{fig4}(a) and Fig.\ref{fig5}, it can be seen that as the QAM modulation order increases, the BER performance of all systems decreases. The reason for this is that higher modulation orders increase the similarity between symbols, making them more difficult to distinguish.

\begin{figure}[htp]
	\centering
	\subfigure{
		{\includegraphics[width=0.3\textwidth]{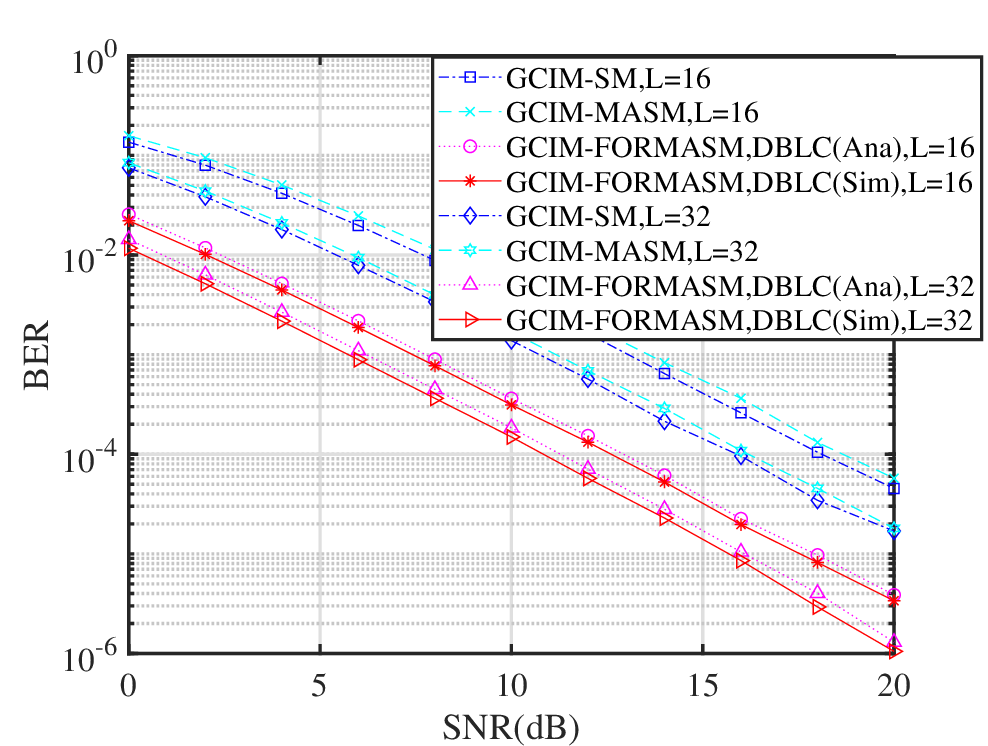}}}
	\caption{Performance comparison for GCIM-SM, GCIM-MASM and GCIM-FORMASM systems, $M=8$, $N_T=4$, $N=3$, $N_R=2$ and $J=8$.  }
	\label{fig6}
\end{figure}
Fig.\ref{fig6} plots the BER performance comparison of various systems related to generalized index modulation under different numbers of Walsh codes. It is undoubtedly true that as the number of Walsh codes increases, the number of data information bits transmitted by all systems also increases. From the figure, it can be observed that with the increase in the number of codes, the BER performance of all systems improves. At the same BER, doubling the number of codes improves the SNR of all systems by approximately 2 dB. However, the gap between the proposed GCIM-FORMASM system and the GCIM-SM, GCIM-MASM systems remains unchanged. Specifically, at the same BER, the GCIM-FORMASM system shows an improvement of approximately 6 dB over the GCIM-SM and GCIM-MASM systems.

\begin{figure}[htp]
	\centering
	\subfigure{
		{\includegraphics[width=0.30\textwidth]{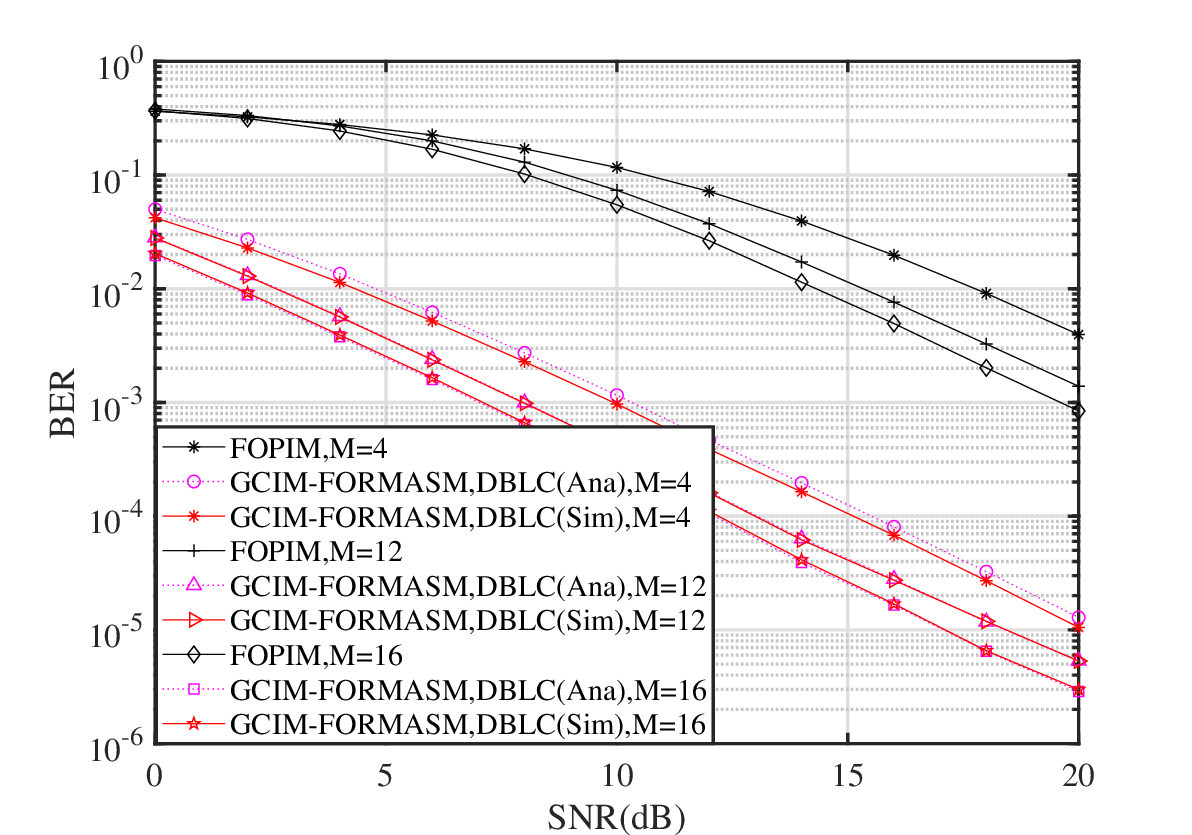}}}
	\caption{Performance comparison for FOPIM and GCIM-FORMASM systems, $L=8$, $N_T=4$, $N=3$, $N_R=2$ and $J=8$. }
	\label{fig7}
\end{figure}

Fig.\ref{fig7} illustrates the BER versus SNR performance for the FOPIM and GCIM-FORMASM systems under various transmission frequency offset pools. The results clearly demonstrate that the proposed GCIM-FORMASM system significantly outperforms the FOPIM system. Notably, at a BER of \(10^{-2}\), the GCIM-FORMASM system requires only about 4 dB, while the FOPIM system necessitates approximately 18 dB. This substantial improvement is evident even when the data transmitted by the two systems differ, underscoring the superior performance of the GCIM-FORMASM system.

\begin{figure}[htp]
	\centering
	\subfigure{
		{\includegraphics[width=0.30\textwidth]{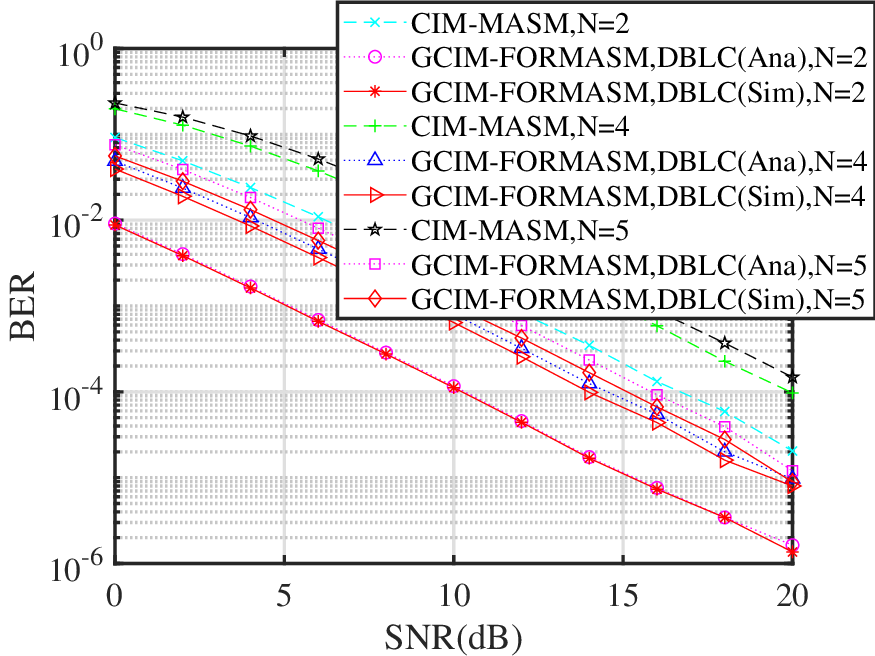}}}
	\caption{Performance comparison for GCIM-MASM and GCIM-FORMASM systems, $L=8$, $N_T=6$, $N_R=2$ and $J=8$. }
	\label{fig8}
\end{figure}

Fig.\ref{fig8} presents the BER performance results of the GCIM-MASM and GCIM-MASM systems with different numbers of activated antennas. It can be observed from the figure that as the number of activated antennas increases, the performance of both systems deteriorates. This is not surprising, as an increase in the number of activated antennas results in a higher number of transmitted information bits.
\section{Conclusion}
\label{sec5}

This paper proposes a GCIM-FORMASM system based on FDA-MIMO, where the information bits are carried not only by conventional modulation methods like QAM but also by spatial index modulation, transmission frequency index modulation, and spreading code index modulation. Subsequently, we utilize the orthogonal waveform transmitted by the FDA to design the corresponding transmitter and receiver structures and provide specific expressions for the received signals. To reduce the decoding complexity of the ML algorithm, we propose a low-complexity DBLC algorithm leveraging the orthogonality of spreading codes. Additionally, this paper presents a closed-form expression for the upper bound of the ABEP of the DBLC algorithm and analyzes its energy efficiency, data rate, and algorithm complexity. Besides, we designed a low-complexity decoding algorithm for GCIM-MASM System. Numerical simulation results validate the correctness of the proposed algorithm, showing that it achieves a lower BER compared to existing algorithms. Moreover, the proposed DBLC algorithm has lower computational complexity and better BER performance compared to the ML algorithm.

\section*{Acknowledgment}
The authors extend their gratitude to the anonymous reviewers for their insightful comments and suggestions, which have significantly enhanced the quality of this paper. Additionally, special thanks are owed to Dr. Jiangwei Jian of the University of Electronic Science and Technology of China for his suggestions during the development of this paper, and to Prof. Guofa Cai of Guangdong University of Technology for his invaluable advice on the coding aspects.
\appendices
\section{Proof for ML Detector}
\label{refA}
Recalling \eqref{eq17}, we can obtain the output of $I$ channel 
\begin{equation}
	\begin{split}
		\tilde{y}_{n_r,k}^{I}=&\sqrt{\frac{P_S}{N}}\sum_{a_n=1}^{a_N}{{\sum_{m=1}^M{h_{\left( \bar{u}_n-1 \right) N+a_n,n_r}x_{\Re}^{a_n,v}{c}_{\Re}^{\left( a_n-1 \right) N_T+i_{\Re},k}}}}\nonumber
		\\
		\,\,      &\times \cos \left[ -2\pi ( f_0+\bigtriangleup f_{u_{\bar{n}},a_n}^{} ) \tau _{a_n} \right] +v_{\left( n_r-1 \right) M+m,k}^{I}.\nonumber
	\end{split}
\end{equation}

Similarly, we have
\begin{equation}
	\begin{split}
		\tilde{y}_{n_r,k}^{Q}=&\sqrt{\frac{P_S}{N}}\sum_{a_n=1}^{a_N}{{\sum_{m=1}^M{h_{\left( \bar{u}_n-1 \right) N+a_n,n_r}x_{\Im}^{a_n,v}{c}_{\Im}^{\left( a_n-1 \right) N_T+i_{\Im},k}}}}\nonumber
		\\
		\,\,      &\times \sin \left[ -2\pi ( f_0+\bigtriangleup f_{u_{\bar{n}},a_n}^{} ) \tau _{a_n} \right] +v_{\left( n_r-1 \right) M+m,k}^{Q}.\nonumber
	\end{split}
\end{equation}
Subsequently, the corresponding channel matrix can be formulated as
\begin{equation}
	\mathring{\boldsymbol{H}}=\boldsymbol{HS}=\big[ \mathring{\boldsymbol{h}}_{1}^{T},\cdots \mathring{\boldsymbol{h}}_{n_r}^{T},\cdots ,\mathring{\boldsymbol{h}}_{N_R}^{T} \big]^{T} \in \mathbb{C} ^{N_R\times N},
\end{equation}
with the selected matrix $\boldsymbol{S}=\left[ \boldsymbol{s}_1,\cdots \boldsymbol{s}_n,\cdots ,\boldsymbol{s}_N \right] \in \mathbb{C} ^{N_TM\times N}$,
where $\boldsymbol{s}_n$ can be expressed as $\boldsymbol{s}_n=\left[ 0,\cdots 1,\cdots ,0 \right] ^T\in \mathbb{C} ^{N_TM\times 1}$.
$\boldsymbol{s}_n$ indicates that only the $
\left( \bar{u}_n-1 \right) N_T+a_n
$th element is non-zero. 

Define 
\begin{equation}
	\boldsymbol{A}=\mathrm{diag}\left\{ \mathfrak{a} _{1}^{},\cdots ,\mathfrak{a} _{n}^{},\cdots ,\mathfrak{a} _{N}^{} \right\} \in \mathbb{C} ^{N\times N},
\end{equation}
with $
\mathfrak{a} _{n}^{}=\cos \left[ -2\pi \left( f_0+\bigtriangleup f_{\bar{u}_n,a_n}^{} \right) \tau _{a_n} \right] +j\sin \left[ -2\pi \left( f_0+\bigtriangleup f_{\bar{u}_n,a_n}^{} \right) \tau _{a_n} \right] 
$. Moreover, $\boldsymbol{A}$ can be divided into the real component $
\boldsymbol{A}^I
$ and imaginary part $
\boldsymbol{A}^Q
$. Denote 
\begin{equation}
	\boldsymbol{X}=\left[ \begin{matrix}
		X_{a_1,1}^{}&		\cdots&		X_{a_1,K}^{}\\
		\vdots&		\ddots&		\vdots\\
		X_{a_N,1}^{}&		\cdots&		X_{a_N,K}^{}\\
	\end{matrix} \right] \in \mathbb{C} ^{N\times K},
\end{equation}
with 
\begin{equation}
	X_{a_n,k}^{}=x_{\Re}^{a_n,v}c_{\Re}^{\left( a_n-1 \right) N_T+i_{\Re},k}+jx_{\Im}^{a_n,v}\boldsymbol{c}_{\Im}^{\left( a_n-1 \right) N_T+i_{\Im},k}.\nonumber
\end{equation}
Further, $\boldsymbol{X}^{I}$ and $\boldsymbol{X}^{Q}$ are the $I$ and $Q$ components of $\boldsymbol{X}^{}$, respectively.

Hence, applying the above results yields 
\begin{equation}
	\begin{split}
	\tilde{\boldsymbol{y}}_{n_r}=&\tilde{\boldsymbol{y}}_{n_r}^{I}+j\tilde{\boldsymbol{y}}_{n_r}^{Q}
	\\
	=&\sqrt{\frac{P_S}{N}}\mathring{\boldsymbol{h}}_{n_r}\left( \boldsymbol{A}^I\boldsymbol{X}^I+j\boldsymbol{A}^Q\boldsymbol{X}^Q \right) +\boldsymbol{v}\in \mathbb{C} ^{1\times K}.\nonumber
	\end{split}
\end{equation}
Next, combining the signal for all receiving antennas results in
\begin{equation}
	\begin{split}
		\tilde{\boldsymbol{Y}}=&\sqrt{\frac{P_S}{N}}\mathring{\boldsymbol{H}}\boldsymbol{AX}+\boldsymbol{V}
		\\
		=&\sqrt{\frac{P_S}{N}}\boldsymbol{HS}\left( \boldsymbol{A}^I\boldsymbol{X}^I+j\boldsymbol{A}^Q\boldsymbol{X}^Q \right) +\boldsymbol{V}
		\\
		=&\sqrt{\frac{P_S}{N}}\boldsymbol{H}\left( \boldsymbol{G}^I\boldsymbol{X}^I+j\boldsymbol{G}^Q\boldsymbol{X}^Q \right) +\boldsymbol{V},
	\end{split}
\end{equation}
with $
\boldsymbol{G}=\boldsymbol{G}^I+j\boldsymbol{G}^Q
$. That means 
\begin{align}
	\boldsymbol{G}^I=&\big[ \boldsymbol{g}_{1}^{I},\cdots \boldsymbol{g}_{n}^{I},\cdots ,\boldsymbol{g}_{N}^{I} \big] \in \mathbb{C} ^{N_TM\times N},\nonumber\\
	\boldsymbol{G}^Q=&\big[ \boldsymbol{g}_{1}^{Q},\cdots \boldsymbol{g}_{n}^{Q},\cdots ,\boldsymbol{g}_{N}^{Q} \big] \in \mathbb{C} ^{N_TM\times N}.\nonumber
\end{align}
with 
\begin{align}
	\label{eq97}
\boldsymbol{g}_{n}^{I}=&\left[ 0,\cdots \cos \big( -2\pi ( f_0+\bigtriangleup f_{\bar{u}_n,a_n}^{} ) \tau _{a_n} \big) ,\cdots ,0 \right] ^T,\\
\label{eq98}
\boldsymbol{g}_{n}^{Q}=&\left[ 0,\cdots \sin \big( -2\pi ( f_0+\bigtriangleup f_{\bar{u}_n,a_n}^{} ) \tau _{a_n} \big) ,\cdots ,0 \right] ^T.
\end{align}
\eqref{eq97}-\eqref{eq98} signify that $\boldsymbol{g}_{n}^{I} $ and $ \boldsymbol{g}_{n}^{Q}$ are all zeros except the $
\left( \bar{u}_n-1 \right) N_T+a_n
$th element.

Based on the ML rule, the estimates of $\boldsymbol{G}$ and $
\boldsymbol{X}
$ can be expressed as 
\begin{equation}
	\label{eq99}
	[ \hat{\boldsymbol{G}},\hat{\boldsymbol{X}} ] =\underset{\boldsymbol{G},\boldsymbol{X}}{arg\,\,\min}\big\| \tilde{\boldsymbol{Y}}-\sqrt{\frac{P_S}{N}}\boldsymbol{H}\left( \boldsymbol{G}^I\boldsymbol{X}^I+j\boldsymbol{G}^Q\boldsymbol{X}^Q \right) \big\| ^2.
\end{equation}

\bibliographystyle{IEEEtran}
\bibliography{ref}
\end{document}